\documentclass[sigconf]{acmart}%

\usepackage{balance}    %
\usepackage{color}
\usepackage{color, colortbl}
\usepackage{array}
  
\usepackage{caption}
\usepackage{subcaption}

\usepackage{multirow}

\usepackage{verbatim}  %

\usepackage{xspace}

\usepackage{graphicx}
\usepackage{xcolor}

\definecolor{Gray}{gray}{0.97}
\definecolor{MedGray}{gray}{0.9}
\definecolor{greytext}{gray}{0.5}
\definecolor{DarkGreen}{rgb}{0.0, 0.5, 0.0}
\definecolor{PFGreen}{rgb}{0.0, 0.5, 0.0}
\definecolor{lightGreen}{rgb}{0.8, 0.9, 0.8}
\definecolor{CadmiumGreen}{rgb}{0.0, 0.42, 0.24}
\definecolor{DarkKhaki}{rgb}{0.74, 0.72, 0.42}
\definecolor{DarkRed}{rgb}{0.7, 0.2, 0.2}
\definecolor{Purple}{rgb}{0.7,0.0,0.7}
\definecolor{Brown}{rgb}{0.7,0.3,0}
\definecolor{Orange}{rgb}{1, 0.5, 0.1}
\definecolor{niceblue}{rgb}{0.0, 0.2, 0.4}

\newcommand{\dissolvpcb}[0]{\textsc{DissolvPCB}\xspace}

\newcommand{\edited}[1]{\textcolor{black}{#1}}
\graphicspath{
{figures/} %
}

\usepackage[nameinlink,capitalise]{cleveref}
\crefname{enumi}{}{}  %
\crefrangeformat{figure}{Figures~#3#1#4--#5#2#6}

\AtBeginDocument{%
  \providecommand\BibTeX{{%
    \normalfont B\kern-0.5em{\scshape i\kern-0.25em b}\kern-0.8em\TeX}}}

\renewcommand\footnotetextcopyrightpermission[1]{%
  \footnotetext[0]{\vspace{10\baselineskip}\hspace{-1em}}
}

\copyrightyear{2025}
\acmYear{2025}
\setcopyright{cc}
\setcctype{by-nc-sa}
\acmConference[UIST '25]{The 38th Annual ACM Symposium on User Interface Software and Technology}{September 28-October 1, 2025}{Busan, Republic of Korea}
\acmBooktitle{The 38th Annual ACM Symposium on User Interface Software and Technology (UIST '25), September 28-October 1, 2025, Busan, Republic of Korea}\acmDOI{10.1145/3746059.3747604}
\acmISBN{979-8-4007-2037-6/2025/09}

\usepackage{siunitx}
\DeclareSIUnit{\mil}{mil}
\DeclareSIUnit{\oz}{oz}
\DeclareSIUnit{\sqft}{ft\textsuperscript{2}}
\usepackage{algorithm2e}
\usepackage{amsmath}

\begin{document}

\title{\dissolvpcb: Fully Recyclable 3D-Printed Electronics with Liquid Metal Conductors and PVA Substrates}

\author{Zeyu Yan}
\email{zeyuy@umd.edu}
\affiliation{%
  \institution{University of Maryland}
  \city{College Park}
  \state{Maryland}
  \country{USA}
} 

\author{SuHwan Hong}
\email{shong999@umd.edu}
\affiliation{%
  \institution{University of Maryland}
  \city{College Park}
  \state{Maryland}
  \country{USA}
} 

\author{Josiah Hester}
\email{josiah@gatech.edu}
\affiliation{%
  \institution{Georgia Institute of Technology}
  \city{Atlanta}
  \state{Georgia}
  \country{USA}
} 

\author{Tingyu Cheng}
\email{tcheng2@nd.edu}
\affiliation{%
  \institution{University of Notre Dame}
  \city{Notre Dame}
  \state{Indiana}
  \country{USA}
} 

\author{Huaishu Peng}
\email{huaishu@umd.edu}
\affiliation{%
  \institution{University of Maryland}
  \city{College Park}
  \state{Maryland}
  \country{USA}
}

\renewcommand{\shortauthors}{Yan, et al.}

\begin{abstract}

We introduce \dissolvpcb, an electronic prototyping technique for fabricating fully recyclable printed circuit board assemblies (PCBAs) using affordable FDM 3D printing, with polyvinyl alcohol (PVA) as a water-soluble substrate and eutectic gallium-indium (EGaIn) as the conductive material. 
When obsolete, the PCBA can be easily recycled by \edited{immersing it} in water: the PVA dissolves, the EGaIn \edited{re-forms into a} liquid metal bead, and the electronic components \edited{are} recovered. \edited{These materials can then be reused to fabricate a new PCBA.} 

We present the \dissolvpcb workflow, characterize its design parameters, evaluate the performance of circuits produced with it, and quantify its environmental impact through a lifecycle assessment (LCA) comparing it to conventional CNC-milled FR-4 boards. 
We further develop a software plugin that automatically converts PCB design files into 3D-printable circuit substrate models. 
To demonstrate the capabilities of \dissolvpcb, we fabricate and recycle three functional prototypes: a Bluetooth speaker featuring a double-sided PCB, a finger fidget toy with a 3D circuit topology, and a shape-changing gripper enabled by Joule-heat-driven 4D printing. 
The paper concludes with a discussion of current technical limitations and opportunities for future directions.
\end{abstract}



\begin{teaserfigure}
    \centering
    \includegraphics[width=\textwidth]{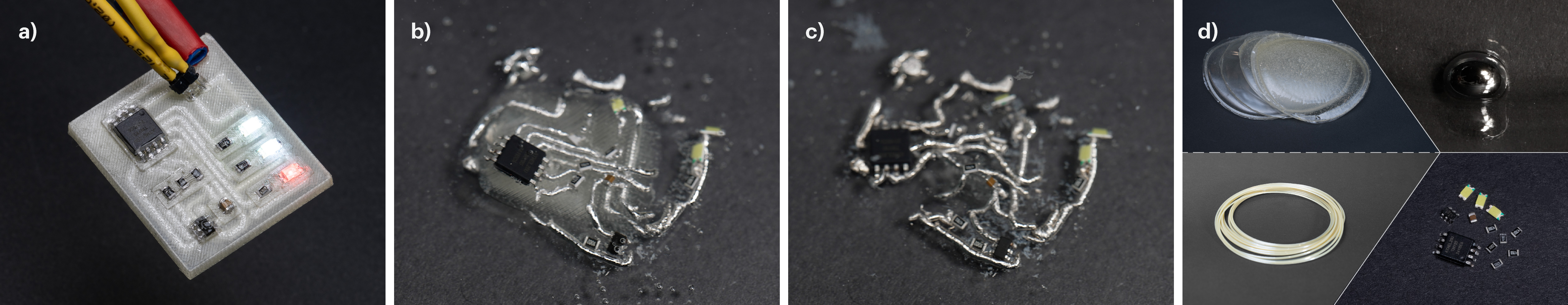}
    \caption{A magnetic field detector circuit made using \dissolvpcb: a) the complete \dissolvpcb assembly; b) the assembly dissolving in water; c) the assembly fully dissolved, with the EGaIn and electronic components separated from the PVA solution; d) dried PVA, re-extruded PVA filament, recovered EGaIn, and retrieved electronic components.}
    \label{fig:teaser}
\end{teaserfigure}

\maketitle

\section{Introduction}

The rapid growth of consumer electronics has been a double-edged sword. 
While the democratization of electronics has modernized nearly every aspect of our lives, it has also led to a significant increase in electronic waste (e-waste). 
According to the annual Global E-Waste Monitor, approximately 62 million metric tons of e-waste are generated globally each year, yet less than 23\% is formally collected and recycled~\cite{balde2024global}. 
As the number of electronic devices continues to grow~\cite{deloitte2025}, traditional e-waste management strategies such as centralized collection and recycling have proven insufficient to meet the scale of the challenge~\cite{balde2024global}.

\edited{Importantly, this problem extends beyond industrial production and mass consumption. Over the past decade, the growing availability of personal fabrication tools and rapid prototyping methods has also contributed to a rise in individual-level e-waste~\cite{yan2023future, song2024sustainable}. This trend underscores the need to rethink how we design, prototype, and manage bespoke electronics, as well as to more fully consider the material and environmental impacts of these practices~\cite{MakeMaking}.}

In response, several technical solutions have been proposed to address e-waste in electronics prototyping. 
For example, ecoEDA~\cite{ecoeda} provides software that facilitates the reuse of electronic components by helping users locate them on obsolete PCBs. 
SolderlessPCB~\cite{solderlesspcb} and PCB Renewal~\cite{yan2025pcbrenewal} promote PCBA recycling through integrated hardware–software ecosystems: the former supports electronic components reuse by eliminating soldering via custom 3D-printed fixtures, while the latter enables substrate reuse by repairing circuit traces with silver epoxy. 
However, these works remain constrained by the traditional FR-4–based PCB fabrication paradigm, and each only supports the partial recovery of a PCBA.

In this paper, we revisit the recycling potential of PCBAs by rethinking the foundations of PCB materials \edited{as well as their manufacturing processes}.
Material-wise, rather than relying on FR-4—the conventional PCB substrate composed of copper-clad fiberglass laminated with epoxy, which is durable but notoriously difficult to recycle—we propose a new PCBA composite that uses a water-dissolvable polyvinyl alcohol (PVA) dielectric and eutectic gallium–indium (EGaIn), a type of liquid metal (LM), for conductors. 
\edited{Manufacturing-wise, instead of lamination, we directly 3D print the main circuit substrate using an off-the-shelf FDM 3D printer, with EGaIn injected into predefined 3D channels to form circuit traces.} We call our approach \dissolvpcb (Figure~\ref{fig:teaser}).

\dissolvpcb functions similarly to traditional FR-4-based PCBAs, as it supports both through-hole (THT) and surface-mounted (SMD) components on either single- or double-sided assemblies. 
What sets it apart is the easy and low-barrier recycling process. 
When no longer needed, a \dissolvpcb assembly can simply be dropped into a tank of water: the PVA substrate dissolves and can be reprocessed into new 3D printing filament, the EGaIn separates into reusable liquid metal droplets, and the electronic components, untouched by solder, can be directly reclaimed (Figure~\ref{fig:teaser}b–d). \edited{A new PCBA prototype, if needed, can then be fabricated from the recycled PVA filament, recovered liquid metal, and reclaimed components.}

\dissolvpcb is also highly accessible and versatile. While previous works have explored alternative PCB substrate materials for sustainable electronics (e.g.,~\cite{teng2019liquid, zhang2023recyclable}), \dissolvpcb is unique for its simple, additive fabrication process and its use of widely available materials that can be sourced from local stores or online vendors (e.g., Amazon). This makes \dissolvpcb potentially scalable and adoptable by many makerspaces. Its additive nature also enables rapid prototyping of circuit designs beyond the capabilities of traditional PCBAs, including 3D artifacts with integrated form and electrical function, or programmable mechanical behaviors enabled by 4D printing.

In the remainder of the paper, we first introduce the design, fabrication, and recycling workflow of \dissolvpcb.  
We then validate its performance and characterize key design parameters through a series of experiments, including 3D-printed trace dimensions, minimum insulation distance between traces and layers, trace conductivity, current capacity, and high frequency waveform transmission performance. 
To integrate our approach into the conventional PCB design pipeline, we developed an open-source FreeCAD~\cite{freecad} plugin\footnote{Open-source \dissolvpcb plugin: \url{https://github.com/zyyan20h/DissolvPCB.git} \label{fn1}} that automatically converts KiCad~\cite{kicad} PCB design files into 3D-printable models for multi-layer PCBs. 
We further assess the sustainability impact of \dissolvpcb by conducting a lifecycle assessment (LCA)~\cite{hauschild2018life} comparing an example circuit to its equivalent manufactured on an FR-4 substrate using a CNC milling machine. 
We demonstrate \dissolvpcb with three fully recyclable examples, a Bluetooth speaker featuring a double-sided PCB, a finger fidget toy with a 3D circuit topology, and a shape-changing gripper enabled by 4D printing. 
We conclude with a discussion of the technical limitations of \dissolvpcb and explore future directions for its integration into personalized PCB design. We also consider \edited{the implications of adopting \dissolvpcb for} mass PCB fabrication, with the \edited{ultimate goal} of addressing the global e-waste challenge.

\section{Related Work}
Our work draws inspiration from prior research in sustainable human-computer interaction (SHCI), the fabrication of sustainable electronics, and, more specifically, studies involving LM and PVA for sustainable manufacturing.

\subsection{SHCI and Sustainable Making} 
The seminal work on Sustainable Interaction Design~\cite{SID} marks decades of research in the field now commonly referred to as sustainable HCI or SHCI. 
While Blevis’s original framework emphasizes the designer’s role and responsibility in integrating sustainability considerations into every phase of a (digital) product or service's design, the field has more recently expanded its focus to include sustainable approaches, practices, and reflections tied to the making, and unmaking of physical artifacts~\cite{song2024sustainable,yan2023future}.

Among many of the ongoing discussions, handling material waste emerges as a recurring theme that draws attention from researchers. For example, Yan et al.~\cite{MakeMaking}, through a qualitative study involving designers, researchers, makers, and makerspace managers, explored how waste materials and electronics are handled and discarded in modern maker environments. They conclude by calling for new tools and infrastructures that support sustainable making practices beyond the traditional centralized recycling paradigm. Similarly, Kim and Paulos~\cite{Creative_Reuse} examined the creative reuse of e-waste through online surveys and observations, through which they proposed a reuse composition framework to guide creative ways of approaching material reuse. The intersection of reuse, recycling, and repair with the material aspects of making has given rise to a family of related concepts such as sustainable unmaking~\cite{unmaking}, uncrafting~\cite{murer2015crafting}, and unfabricating~\cite{Unfabricate}, all centered on exploring the afterlife of obsolete designs, products, or physical forms.

New technical approaches have also been proposed in SHCI, with researchers exploring both novel fabrication processes and sustainable material alternatives. One area of focus has been the development of computational tools to address material waste in 3D printing. These include tools to repurpose obsolete prints as infill for new designs~\cite{Scrappy}, reduce the use of support materials~\cite{Substiports}, maximize the use of leftover filament~\cite{deshpandeunmake}, and support the recycling of multi-material 3D prints~\cite{wen2025recycling}.

Other research efforts have explored sustainable making through the use of novel materials. For example, to promote sustainability in textiles, projects such as EcoThreads~\cite{ecothread} and Desktop Biofiber Spinning~\cite{spinning} have investigated water-dissolvable yarns for rapid textile prototyping. To reduce the use of thermoplastics in 3D printing, recent studies have proposed using waste coffee grounds~\cite{Coffee_Grounds} or compostable play-dough~\cite{Printable_Play-Dough} as alternative 3D printing materials.
Organic or bio-based materials have also been explored for prototyping physical artifacts, including yeast~\cite{SCOBY}, fungi~\cite{Myco-Accessories}, algae~\cite{Alganyl}, chitosan~\cite{song2022towards}, and cellulose-based fibers~\cite{Cellulose-Based_Optical_Textile_Sensors, Bioplastics}.

Our work aligns with prior literature in aiming to reduce material waste in sustainable making through the use of novel materials and computational tools. 
However, we specifically focus on sustainable electronics and PCBA fabrication.

\subsection{Sustainable Electronics} 
As mentioned earlier, most PCBAs in modern electronic devices are made using FR-4, a composite of woven fiberglass cloth and epoxy resin laminated with copper foil on one or both sides. 
When these devices reach end-of-life, they contribute to global e-waste. Reuse is rare due to product-specific designs, and recycling remains difficult because the materials are tightly bonded and hard to separate.

The primary approach in today's electronics management industry is to partially recover selective high-value materials from PCBAs using centralized recycling facilities, where PCBAs undergo mechanical and chemical treatments that break down them into undifferentiated scrap for material recovery~\cite{su131810357, Kiddee2013}. 
This centralized process has several drawbacks. 
It requires large-scale transportation infrastructure to collect and ship PCBA waste, overlooks the potential for reusing many functional electronic components, and the recycling process itself can produce toxic fumes, is energy-intensive, and may cause additional environmental damage~\cite{cui2003mechanical}.

There are mainly two threads of research that aim at a more sustainable approach to reduce e-waste. 
One group of HCI literature focuses on turning the existing FR-4 centered PCB infrastructure more sustainable.  
For example, PCB Renewal~\cite{yan2025pcbrenewal} and ProtoPCB~\cite{ProtoPCB} focus on the FR-4 substrate, proposing new repair techniques and computational systems to repurpose obsolete substrates into new designs. 
ecoEDA~\cite{ecoeda},  SolderlessPCB~\cite{solderlesspcb} and ProForm~\cite{proform} address the reuse of electronic components from PCB assemblies. 
ecoEDA, through a custom electronic design automation (EDA) plugin, provides designers with suggestions for sorting components from old assemblies, while SolderlessPCB and ProForm address the practical challenges of desoldering by introducing solder-free mechanisms that greatly simplify the disassembly of PCBAs.

\begingroup
\sloppy
On the other hand, ongoing efforts are investigating PCB substrates that are inherently easier to recycle. 
For example, paper-based boards~\cite{Silver_Tape, Instant_Inkjet_Circuits, Printed_Paper_Actuator}, wood-derived composites~\cite{Wooden_Circuit, A_Tale_of_Two_Mice}, bio-based materials\edited{~\cite{bharath2020novel, yedrissov2022new, song2022towards, 10.1145/3341162.3343808, guna2016plant}}, and water-soluble options~\cite{cheng2024recy} have been investigated as circuit substrates. 
Transesterification-based vitrimers~\cite{zhang2024recyclable, Biswal2025} and sodium hydroxide (NaOH)–embedded paraffin composites~\cite{teng2023fully} have also been proposed as alternatives to traditional FR-4. 
Among these, Jiva~\cite{jiva} now offers a commercialized, biodegradable substrate that is suitable for scalable production.
The ease with which these alternative substrates break down or can be separated from components supports the broader vision of transient electronic devices~\cite{Transient_Electronics, Transient_PCB}—systems intentionally engineered to simplify material recovery and recycling once the devices reach the end of their lifespan~\cite{cheng2023functional, song2023vim}. 
\endgroup

\dissolvpcb also contributes to the growing body of work focused on developing sustainable PCBAs.
Unlike prior efforts that primarily explore material properties or produce one-off demonstrative examples, our work presents an end-to-end pipeline for fabricating sustainable PCBAs.
This pipeline leverages the existing infrastructure of today’s makerspaces, requiring only an off-the-shelf, unmodified FDM 3D printer and easy-to-source materials.

\subsection{LM and PVA for Sustainable Fabrication}
The two materials used in \dissolvpcb are EGaIn, a type of LM, and PVA in the form of 3D printing filament. This section provides a brief background on both materials, including their properties and roles in sustainable fabrication.

LMs are unique in their combination of high electrical conductivity, low viscosity, and intrinsic stretchability. 
These properties enable LM traces to be patterned on elastomeric substrates and, as an alternative to rigid conductors, allow circuits to bend, stretch, or twist without losing functionality~\cite{dickey2017stretchable, fassler2015liquid, Transient_Electronics, Silicone_Devices}. 
LM has thus been used in applications such as soft robotics, wearable sensors, tangible interfaces, and biomedical devices~\cite{park2010hyperelastic, Flowcuits}. 
In addition, due to their low melting point, high surface tension, and liquid state at room temperature, the fluidity of LMs has been leveraged to create recyclable conductive traces in PCBs~\edited{~\cite{10.1145/3341162.3343808, cheng2024recy}}.

PVA is a water-soluble polymer that can be dissolved on demand and, as a thermoplastic, exhibits a softening point above \SI{180}{\celsius}. 
This recoverability has attracted great attention from researchers in materials science, electronics, as well as HCI, where several studies in these fields have proposed using PVA sheets as substrates for creating transient circuit examples~\cite{cheng2023functional, teng2019liquid}. 
While these studies show the promise of PVA as a circuit substrate, the fabrication of these circuits either involves heavy manual effort or relies on specialized ink-based direct writing hardware onto PVA sheets, which is only suitable for single-sided circuit design. 

Like prior work, \dissolvpcb also adopts the combination of LM and PVA as the main PCB materials. 
However, unlike previous approaches that use premade PVA sheets, our substrate is fabricated through FDM 3D printing using PVA filament spools. 
This fabrication method offers greater flexibility in conductor routing and supports single-, double- and even three-dimensional circuit traces, greatly expanding the capabilities of PVA-based sustainable circuits. 
The accompanying software plugin further enables end users to convert traditionally multi-layer circuit schematics directly into 3D-printable files, lowering the technical barrier to fabrication.

\section{\dissolvpcb}\label{overview}

\dissolvpcb is an enabling PCBA manufacturing technique that combines FDM 3D printing and sustainable materials to create fully recyclable circuit assemblies. 
The method employs PVA filament, a biodegradable polymer traditionally used as a support material in FDM 3D printing, as the primary dielectric, and EGaIn, a well-studied, easily recyclable conductive alloy, as the electrical conductor.
At the end of the PCBA's lifecycle, the entire assembly can be immersed in water for recycling.

\begin{figure}[t]
    \centering
    \includegraphics[width=\columnwidth]{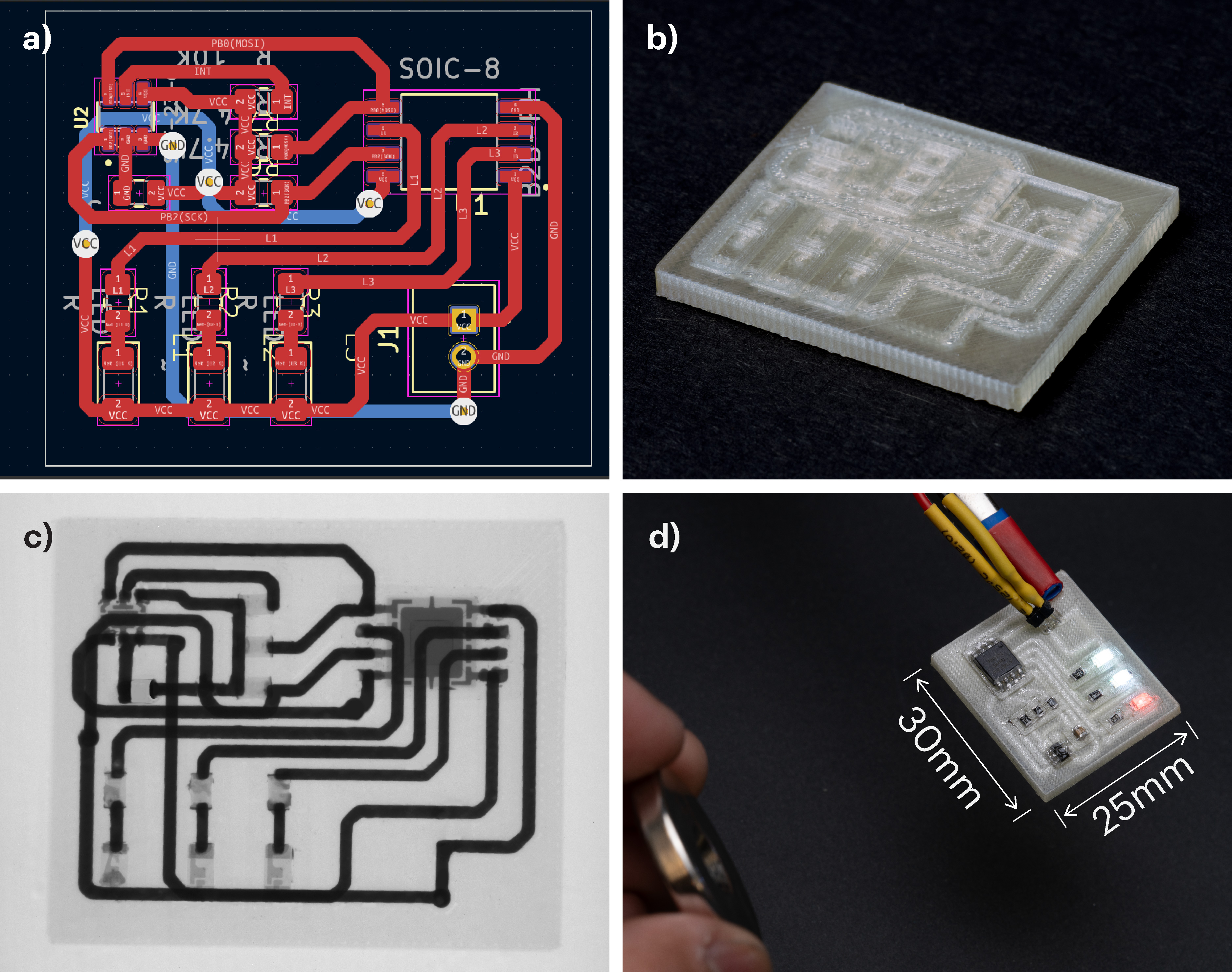}
    \caption{Overview of the magnetic field detector circuit: a) PCB design created in KiCad; \edited{b) 3D-printed substrate of the detector; c) micro-CT scan of the substrate with inserted EGaIn; d) detector activated by a magnet.}}
    \label{fig:benchmark}
\end{figure}

\begin{figure*}[h]
    \centering
    \includegraphics[width=\textwidth]{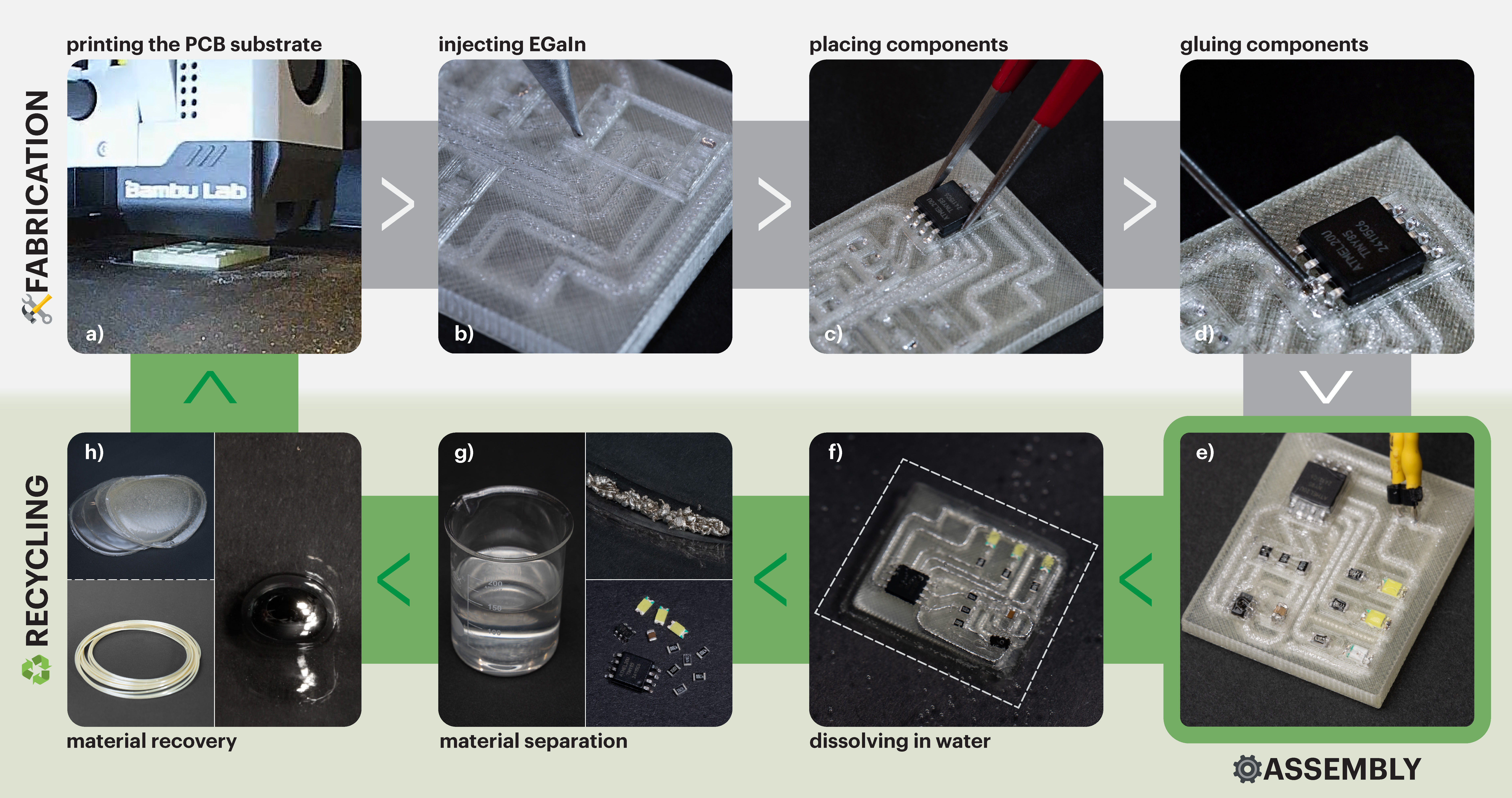}
    \caption{Fabrication and recycling process of \dissolvpcb: a) 3D printing the circuit substrate; b) injecting EGaIn; c) placing components into their corresponding sockets; d) fixing components and sealing the EGaIn channel openings using PVA adhesive; e) completed and functional \dissolvpcb circuit; f) the example circuit dissolving in water; g) separated elements after dissolving, including PVA solution, EGaIn, and electronic components; h) recovered raw materials, including dried PVA sheets, re-extruded PVA filament, and processed EGaIn.}
    \label{fig:fab}
\end{figure*}

In this section, we provide detailed accounts of both the fabrication and recycling processes of PCBAs made using the \dissolvpcb method.
We use a small magnetic field detector circuit shown in Figure \ref{fig:benchmark} as a working example.
This circuit includes an ATtiny85 microcontroller, an $I^2C$ Hall effect sensor, three 1206 LEDs, and seven 0805-package components (six resistors and one capacitor), forming the peripheral circuitry for magnetic field detection. The complete fabrication and recycling process is illustrated in Figure~\ref{fig:fab}.

\subsection{Fabrication of a \dissolvpcb} \label{fab}
Fabricating a recyclable \dissolvpcb PCBA  can be divided into the following steps: 1) creating a 3D-printable PCB substrate, 2) printing the substrate, 3) injecting EGaIn, and 4) inserting and securing the electronic components.

\subsubsection{Creating a 3D-printable PCB substrate.} 
A basic printable PCB substrate consists of three design components: the overall substrate body, which can take the form of a traditional flat panel or a three-dimensional shape; the conductive traces, which are networks of interconnected hollow channels with outlets that extend to the substrate surface; and the component sockets, which are sized to match the electronic components so they can be easily inserted. 
The metal connectors of each electronic component should align with the outlets of the hollow channels to ensure electrical contact. 

Designing such a PCB substrate can be done directly using CAD modeling software, as demonstrated in the examples in Section~\ref{fidget} and Section~\ref{gripper}. 
However, manually routing the hollow channels without an Electrical Rule Check (ERC) increases the likelihood of design errors and requires design expertise. 
To simplify this process, we provide a custom FreeCAD plugin (see Footnote~\ref{fn1}) that automates the substrate design. 
The plugin converts a conventional single- or double-sided PCB design file into a 3D model, translating the conductive trace network into the hollow channels and generating the appropriate sockets. 
Our working example in Figure~\ref{fig:benchmark}, along with the circuit described in Section~\ref{speaker}, were both created using this auto-conversion approach. Details of the software implementation are provided in Section~\ref{software}.

\subsubsection{Printing the substrate}
Unless otherwise noted, all examples in this paper are printed using an off-the-shelf BambuLab P1S FDM printer with a factory-installed \SI{0.2}{\milli\meter} nozzle, at a layer thickness of \SI{0.06}{\milli\meter} and a wall thickness of \SI{0.15}{\milli\meter} (Figure~\ref{fig:fab}a). 
Alternatively, depending on the model of the substrate, a \SI{0.4}{\milli\meter} nozzle can also be used for faster printing. 
Through empirical experiments, we found that a layer thickness of \SI{0.12}{\milli\meter} and a wall thickness of \SI{0.3}{\milli\meter} produced good results with a \SI{0.4}{\milli\meter} nozzle.

While printing the substrate itself is straightforward (and therefore easily scalable), one challenge we encountered during experiments was ensuring clear and reliable channels throughout the substrate. 
A common issue when printing with PVA is the stringing effect, where thin strands of material are left between parts of the print. 
When this occurs inside narrow channels, it can cause blockages that hinder EGaIn injection.

To mitigate this, we empirically set the printing speed to \SI{30}{\milli\meter\per\second} and the retraction distance to \SI{10}{\milli\meter}. Additionally, the ceiling layer that seals the channels was printed using bridge settings at \SI{15}{\milli\meter\per\second}, with 100\% part-cooling fan speed to ensure rapid solidification of the extruded material and to prevent dripping. 
Since conventional PCBs use vertical, horizontal, and \SI{45}{\degree} trace directions, we also set the bridge infill angle to \SI{22.5}{\degree} to accommodate multi-directional traces, thereby avoiding long bridges that could also cause dripping and clogging. 

With the above settings, we can print hollow traces with a minimum cross-section of \SI{0.7}{\milli\meter} by \SI{0.7}{\milli\meter}. 
Section~\ref{evaluation} provides experimental details.

\subsubsection{EGaIn injection}
The EGaIn used in this research was prepared in our lab using raw materials sourced from online vendors such as Amazon.
\edited{While EGaIn is non-toxic, exhibits no vapor pressure at room temperature, and is generally considered safe for human handling, standard laboratory safety practices are still recommended. 
We advise the use of gloves and safety goggles during handling. 
A brief overview of our preparation method is provided below. More detailed instructions are available in Appendix~\ref{appendix} and in the accompanying GitHub repository (see Footnote~\ref{fn1}).}

The preparation of EGaIn requires 75.5\% gallium and 24.5\% indium by weight. 
Since the melting point of gallium is around \SI{29.8}{\degreeCelsius}, we first melt the gallium using a warm water bath. 
Once fully liquefied, indium is gradually added and thoroughly stirred using a non-reactive stirrer. 
The finished EGaIn is stored in a sealed glass container to minimize oxidation.

EGaIn is injected using a small-diameter syringe fitted with tapered blunt needles (25 gauge). 
The injection can begin at any opening of the channel and continues until all channels are filled (Figure~\ref{fig:fab}b). 
For optimal connection performance, EGaIn should ideally form a convex meniscus at each channel opening. 
Although bubble entrapment in the channels was initially a concern, \edited{an X-Ray microtomography (micro-CT) scan  of the sample circuit (Figure \ref{fig:benchmark}c) confirmed that our injection process does not result in significant trapped air.}

\subsubsection{Inserting and securing electronic components}
With EGaIn injected, electronic components can be inserted into their corresponding substrate sockets to complete the circuit (Figure~\ref{fig:fab}c). 
While several previous studies (e.g.,~\cite{cheng2023functional, cheng2024recy}) have demonstrated that the surface tension of EGaIn is sufficient to hold components in place, we introduce an additional step: applying PVA glue at each connection interface  (Figure~\ref{fig:fab}d). 
This step is analogous to soldering electronic components to metal leads, ensuring firm and reliable electrical and mechanical bonding. 
It also fully seals the EGaIn within the channels, preventing it from leaking or oxidizing.

Note that, in theory, any PVA-based glue, such as school glue~\cite{ElmersPVA}, can function similarly in this step. 
However, through our experimentation, we found that formulations with excessively low viscosity risk dissolving the PVA substrate, potentially causing irreversible structural damage or short circuits between traces.
To address this, we prepared a custom mixture of PVA pellets and water at a 3:5 weight ratio, deliberately reducing the water content to minimize substrate dissolution. 
We also selected PVA with a moderate molecular weight range of 31,000–50,000. 
For material preparation, the PVA solution is stirred at \SI{80}{\degreeCelsius} for 2-3 hours, until the pellets are fully dissolved and the glue is ready for use. 

After applying the glue to the component–substrate seams, the assembly is dried in a heated chamber at \SI{60}{\degreeCelsius}.
With our formulation, drying takes approximately one hour, resulting in a bond strength sufficient for normal handling. 
Across all examples and samples produced during our research, we observed no visible signs of electrical failure, such as EGaIn leakage or short circuits, after the glue had dried.

\subsection{Recycling of a \dissolvpcb}\label{recycle}
The PCBA produced through the \dissolvpcb process is straightforward to recycle, as it requires no specialized equipment or harsh chemicals. This simplicity enables recycling in most electronics labs or makerspaces. In addition, all materials used in the \dissolvpcb PCBA, including the PVA substrate, EGaIn traces, and electronic components, can be effectively recovered and reused. Below, we describe the recycling process. A detailed analysis of the environmental impact of \dissolvpcb is provided in Section~\ref{lca}.

\subsubsection{Dissolving the assembly} The recycling process begins by placing the assembly in water. 
As the PVA substrate absorbs water, it gradually dissolves (Figure~\ref{fig:fab}f). 
The time required to dissolve the substrate varies, primarily depending on the amount of energy and resources the user is willing to invest in the process. 
For example, the sample PCBA shown in Figure~\ref{fig:fab}e takes approximately 36 hours to fully dissolve when placed in a small petri dish with just enough water to cover its top, kept at room temperature (\SI{22}{\degreeCelsius}) without heating or stirring. 
In contrast, with active stirring and heating to \SI{90}{\degreeCelsius}, most of the sample dissolves in less than one hour. 
Therefore, depending on the user's preferences and constraints, such as urgency, energy consumption, or available equipment, the dissolving process can be optimized for either convenience or efficiency.

\subsubsection{Recycling electronic components and EGaIn}
After dissolution, the electronic components and EGaIn typically settle at the bottom of the container, naturally separating from each other. The components can be directly retrieved and are ready for reuse after drying, while the residual EGaIn is deoxidized using a minimal amount of NaOH solution, \edited{applied at the droplet scale at \SI{2}{\mole\per\liter}. The resulting small volume of high-pH solution can be neutralized either by adding a dilute solution of citric acid until the pH stabilizes between 6 and 8, or by allowing it to sit exposed to air, where it gradually converts into a mixture of sodium carbonate and sodium bicarbonate.}
This process restores the surface tension of the EGaIn, allowing it to reform into a single bead (Figure \ref{fig:fab}h), which can then be easily extracted from the container using a syringe or pipette.

\subsubsection{Recycling PVA}
The PVA solution can be dried at room temperature or on a hot plate heated to just below the boiling point of water.
Once dried, the PVA can be peeled off from its container in the form of sheets and reused in various ways. For example, it can be re-dissolved to make PVA glue, or shredded and re-extruded into filament using a filament extruder.

\begin{figure}[t]
    \centering
    \includegraphics[width=\columnwidth]{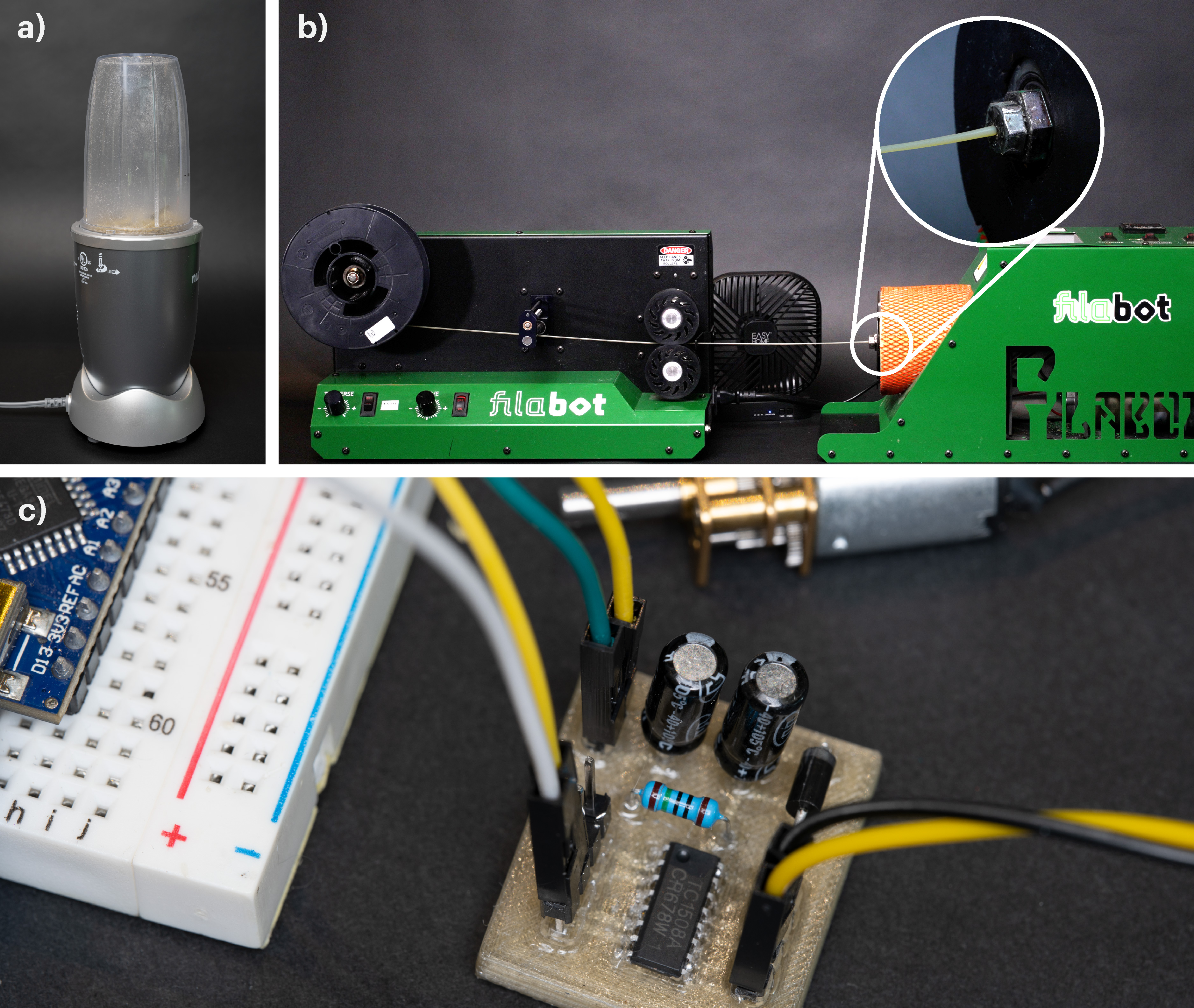}
    \caption{PVA re-extrusion and reuse: a) a shredder used for processing dried PVA; b) re-extruding PVA filament using a filament extruder; c) \dissolvpcb motor driver fabricated with the re-extruded PVA filament.}
    \label{fig:re-ext}
\end{figure}

\edited{Figure~\ref{fig:re-ext}a and b illustrates our process of turning the dried PVA sheets collected throughout this project into new PVA filament. 
Specifically, we ground 16 pieces of dried PVA sheets %
into small chips using a kitchen shredder, which were then fed into a Filabot EX2 filament extruder~\cite{filabotEX2} for re-extrusion.
We measured the output filament diameter every \SI{2.5}{\meter}, and the average diameter was \SI{1.792}{\milli\meter} with a standard deviation of \SI{0.057}{\milli\meter}.}

\edited{This spool of filament has been used both as a 3D printing support material and to fabricate new \dissolvpcb circuits. As one example, Figure~\ref{fig:re-ext}c shows a motor driver module~\cite{MX1508MotorDriverModule} printed using the recycled PVA filament.
The sample features an MX1508 SMD IC, four THT components, and ten THT headers, and it successfully handles the PWM signal used to drive a DC motor.}

\section{Characterization and Technical Evaluation}\label{evaluation}
Since \dissolvpcb\ uses materials and manufacturing processes that differ from traditional FR-4, we conducted a series of experiments to characterize its key design parameters and evaluate the electrical performance of the printed circuits. The results of these design decisions and experiments are presented in this section.

\subsection{Component Socket Design and Affixing} \label{socket}

\begin{figure}[b]
    \centering
    \includegraphics[width=\columnwidth]{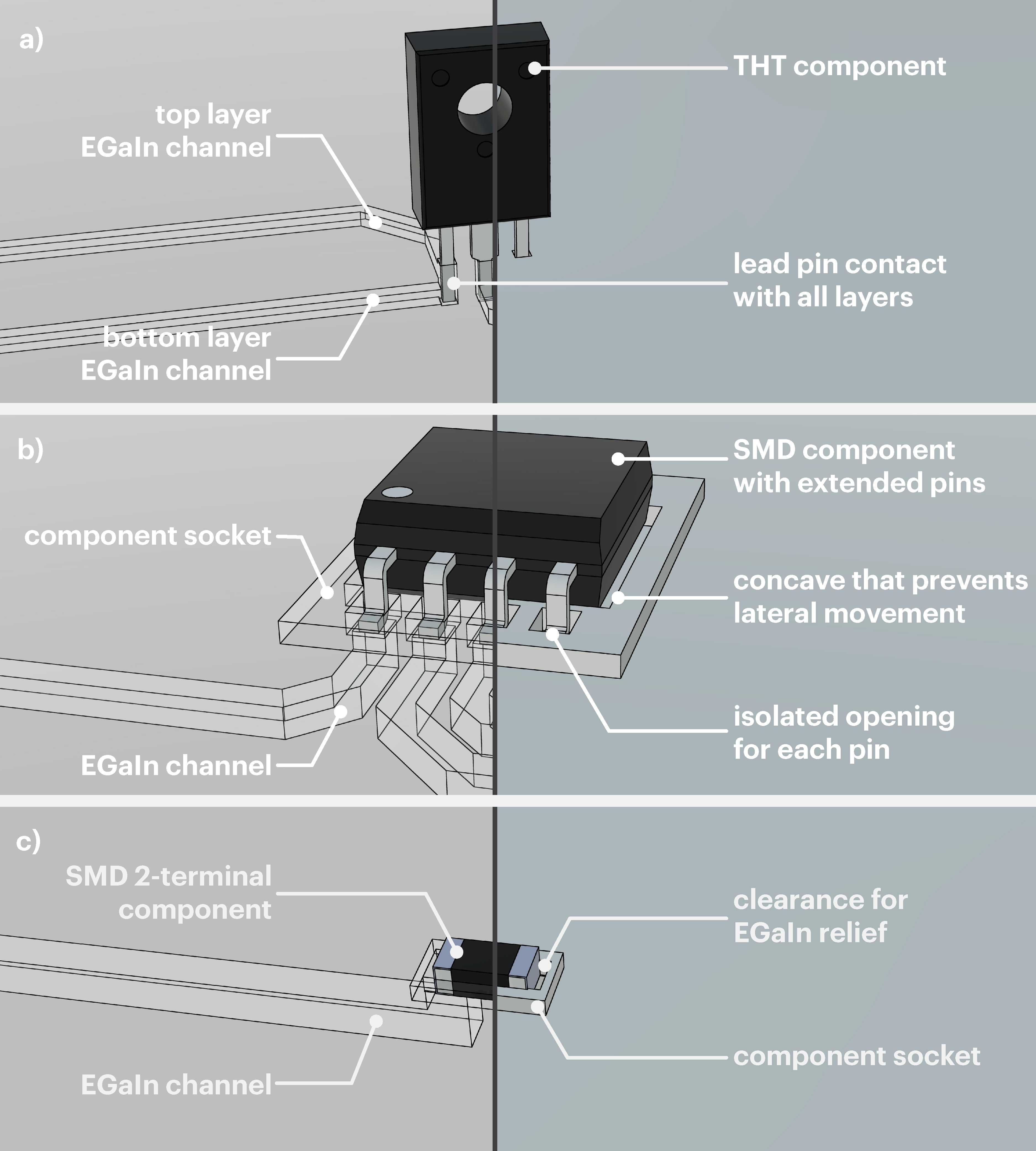}
    \caption{Socket designs for components: a) THT; b) SMD with extended pins; c) two-terminal SMD.}
    \label{fig:socket}
\end{figure}

\edited{As shown in Figure~\ref{fig:socket}, we considered a range of electronic socket designs to accommodate components in various packaging formats. These designs account for the practical printing resolution of our FDM printer and support both THT and SMD components, including those with extended pins and a pitch of \SI{0.85}{\milli\meter} or greater, as well as two-terminal components in the 0603 package or larger. Dedicated sockets were created for each component type to ensure mechanical stability and reliable electrical contact.} 

\edited{Specifically, for THT components (Figure \ref{fig:socket}a), cylindrical or cuboid negative spaces are allocated for each pin, extending through the full depth of the PCB across all layers.
Unlike conventional THT vias in FR-4 PCBs, our design exposes an opening only on the side where the component is inserted. 
This prevents EGaIn leakage during the application and curing of the PVA glue.
After injecting EGaIn into the channels, the component leads can be trimmed to length and directly inserted into the sockets. 
PVA glue is then applied to the single accessible opening at each channel outlet to secure the component.}

\edited{For SMD components with extended leaded pins (Figure \ref{fig:socket}b), we have designed custom sockets that isolate each pin to prevent EGaIn overflow and shorting between adjacent leads prior to sealing. 
Each socket features a shallow concavity to cradle the chip body, minimizing lateral movement during assembly.
PVA glue is applied at the leaded pins; its fluidity allows it to flow over and encapsulate the EGaIn–pin interface, ensuring mechanical stability and electrical insulation once dried.
Two-terminal components (Figure \ref{fig:socket}c) are press-fit into cuboid sockets, each designed with a \SI{0.2}{\milli\meter} clearance at both ends. This clearance directs liquid metal away from adjacent pads, reducing the risk of electrical shorts.}

\edited{Pre-designed 3D models of these component sockets are included in the \dissolvpcb software package (see Section~\ref{software}) as a KiCad library. Users can incorporate them during the PCB design process, just as they would with any standard KiCad component library.}

\subsection{Trace Channel Dimension}
Conductive traces in \dissolvpcb are formed using 3D-printed channel networks filled with EGaIn. While these printed channels may not match the thinness of traces fabricated on an FR-4 substrate, they can still achieve relatively compact dimensions.

\begin{figure}[b]
    \centering
    \includegraphics[width = \columnwidth]{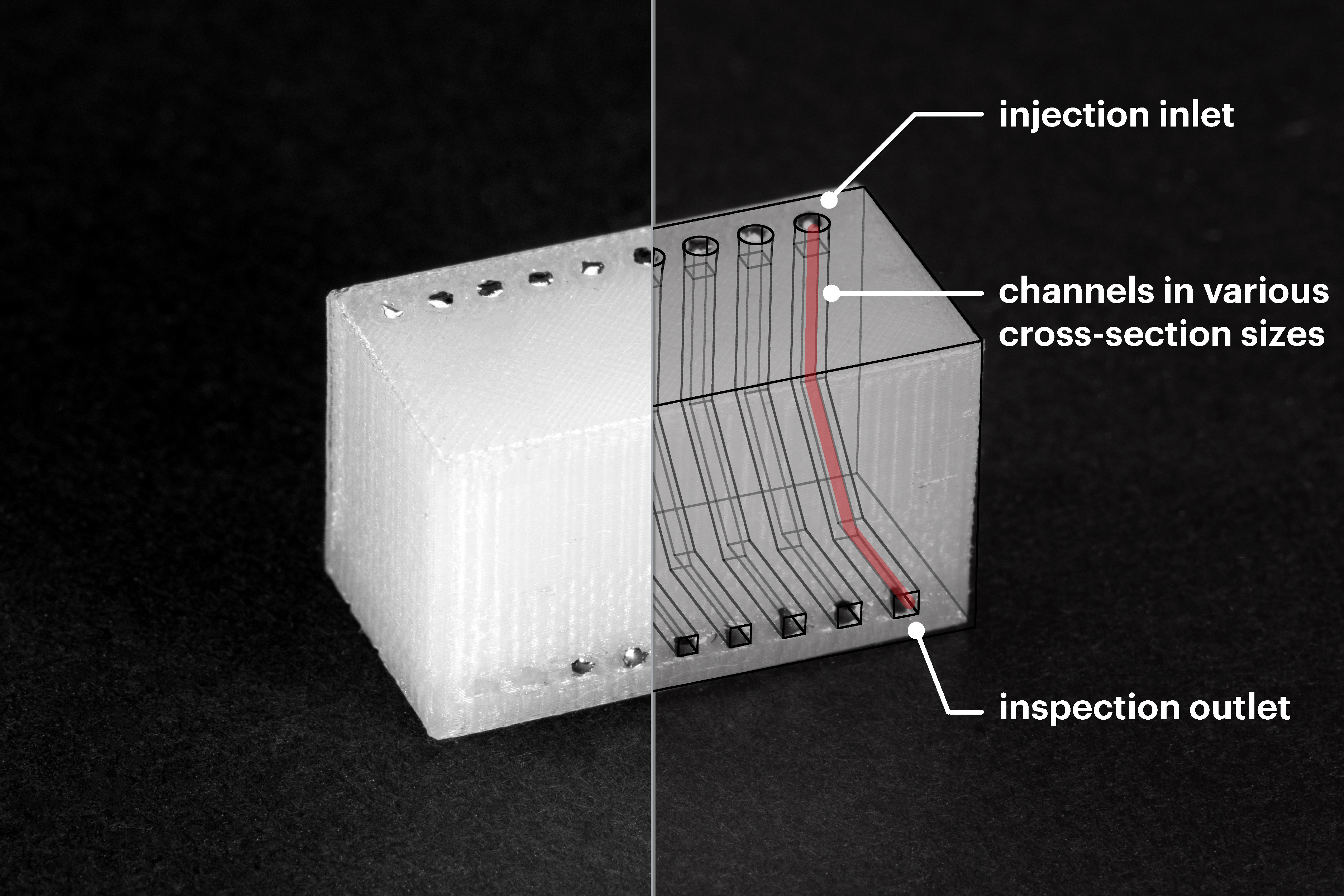}
    \caption{Trace dimension experiment sample and its internal structure, shown on the right.}
    \label{fig:dimension}
\end{figure}

The minimum channel width and height are primarily determined by the 3D printer's nozzle diameter and printing layer height.
To identify the smallest viable channel cross-section dimensions, we fabricated a series of test channels with cross-sections ranging from \SI{0.6}{\milli\meter} to \SI{1.0}{\milli\meter} in \SI{0.05}{\milli\meter} increments (Figure \ref{fig:dimension})\edited{, with three samples per thickness}. 
Each channel included vertical, horizontal, and \SI{45}{\degree} sloped segments.
We then performed EGaIn injection trials to evaluate the filling success rate and the quality of the resulting conductive traces.
\edited{Successful injections were confirmed using a digital multimeter.}

Results showed that channels with cross-sections of \SI{0.7}{\milli\meter} $\times$ \SI{0.7}{\milli\meter} could be reliably filled when printed with a \SI{0.2}{\milli\meter} nozzle. 
With a \SI{0.4}{\milli\meter} nozzle, channels with dimensions of \SI{0.9}{\milli\meter} $\times$ \SI{0.9}{\milli\meter} were reliably produced.
These outcomes were consistently observed across multiple trials.

\subsection{Minimal Insulation Thickness}

To enable circuit designs without malfunction, ensuring dielectric insulation between individual signal nets is essential. 
In \dissolvpcb, dielectric insulation is achieved with PVA material positioned between adjacent channels.
Due to the layered nature of the FDM printing process, the minimum achievable insulation thickness is primarily constrained by the smallest printable wall thickness in either the X/Y plane or along the Z-axis.

\begin{figure}[b]
    \centering
    \includegraphics[width = \columnwidth]{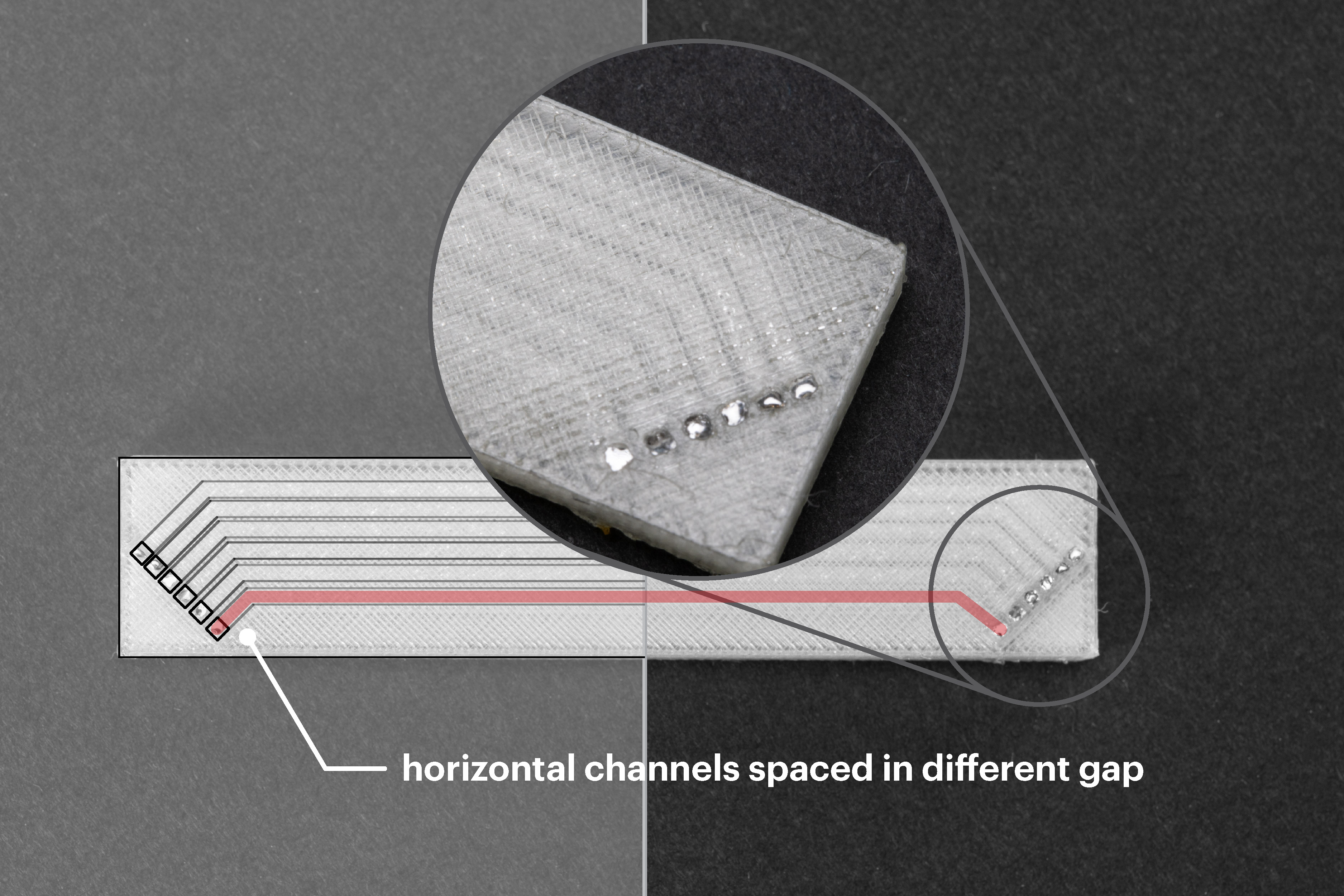}
    \caption{Trace X/Y-direction insulation test sample, with an illustration of the internal structure on the left.}
    \label{fig:XYinsulation}
\end{figure}

To identify the minimum printed wall thickness capable of reliably insulating EGaIn-filled channels, we fabricated test traces with varying wall thicknesses. 
These ranged from the nozzle’s lowest recommended thickness of \SI{0.15}{\milli\meter} for a \SI{0.2}{\milli\meter} nozzle to \SI{0.35}{\milli\meter}, incremented by \SI{0.05}{\milli\meter} (Figure \ref{fig:XYinsulation}). 
No short circuits were observed between any traces, even at the minimum wall thickness of \SI{0.15}{\milli\meter}, across multiple samples.
Combined with the minimum channel cross-section of \SI{0.7}{\milli\meter} $\times$ \SI{0.7}{\milli\meter}, this enables the design of complex circuits compatible with electronic components having pin-to-pin pitch distances equal to or greater than \SI{0.85}{\milli\meter}. 
This includes common THT components such as DIP, SIP, and PGA packages, as well as select SMD packages including SOIC, SSOP, QFP, SOT-23, and PLCC.

\begin{figure}[b]
    \centering
    \includegraphics[width = \columnwidth]{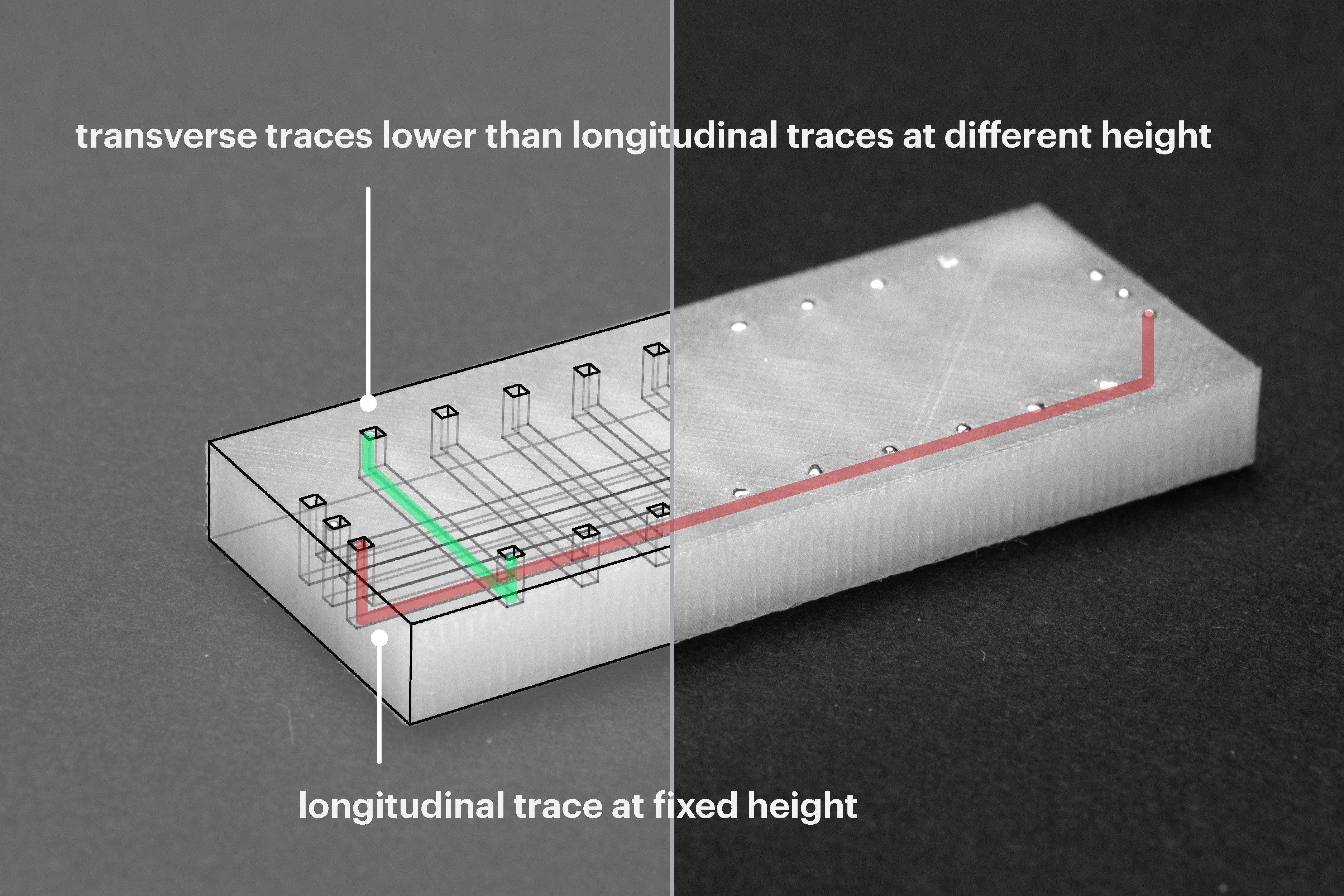}
    \caption{Trace Z-direction insulation test sample, with an illustration of the internal structure on the left.}
    \label{fig:ZInsulation}
\end{figure}

For Z-direction insulation, we printed samples containing three parallel longitudinal traces at the same height and a series of transverse traces at varying heights.
The Z-axis insulation layer thicknesses between the transverse and longitudinal traces ranged from \SI{0.18}{\milli\meter} to \SI{0.66}{\milli\meter} in \SI{0.06}{\milli\meter} increments, corresponding to three to eleven printed layers (Figure \ref{fig:ZInsulation}). 
Each longitudinal-transverse trace pair was tested for short circuits using a digital multimeter. 
We confirmed that traces insulated by three layers (or \SI{0.18}{\milli\meter}) remained fully isolated, establishing this as the minimum  Z-direction insulation thickness.
The combination of \SI{0.7}{\milli\meter} trace thickness and \SI{0.18}{\milli\meter} Z-direction insulation results in a minimum thickness of \SI{1.06}{\milli\meter} for a single-layer PCB substrate and \SI{1.94}{\milli\meter} for a double-layer PCB substrate.
As the 3D printing process can accommodate PCBs with more than two conductive layers, the total thickness scales linearly with the number of layers in the PCB substrate design.

\subsection{Conductivity and Current Capacity}

The conductivity of EGaIn traces is fundamentally determined by the material's volume resistivity. 
Previous research has shown that EGaIn exhibits a resistivity of \SI{29}{\micro\ohm\centi\meter}, approximately ten times higher than that of copper\cite{dickey2008eutectic}. 
In practice, trace conductivity also depends on the trace's cross-sectional metallic area. 
To evaluate the conductivity and current-carrying capacity of EGaIn traces formed within 3D-printed PVA channels, we fabricated \edited{three} \SI{30}{\milli\meter}-long traces with cross-sectional dimensions of \SI{0.7}{\milli\meter} by \SI{0.7}{\milli\meter}. 
Resistance measurements using a multimeter yielded an average value of \SI{0.03}{\ohm} per \SI{30}{\milli\meter} trace length, \edited{with a measured standard deviation of \SI{0.0012}{\ohm}}. 
For comparison, conventional PCBs using 1 oz copper exhibit a resistance of approximately \SI{0.02}{\ohm} per \SI{30}{\milli\meter} trace length at the same width of \SI{0.7}{\milli\meter}, suggesting that EGaIn traces achieve comparable conductivity despite differences in thickness. 
The low resistance of EGaIn traces makes them suitable for many low-voltage, low-current DC circuit applications.

\begin{figure}[t]
    \centering
    \includegraphics[width = \columnwidth]{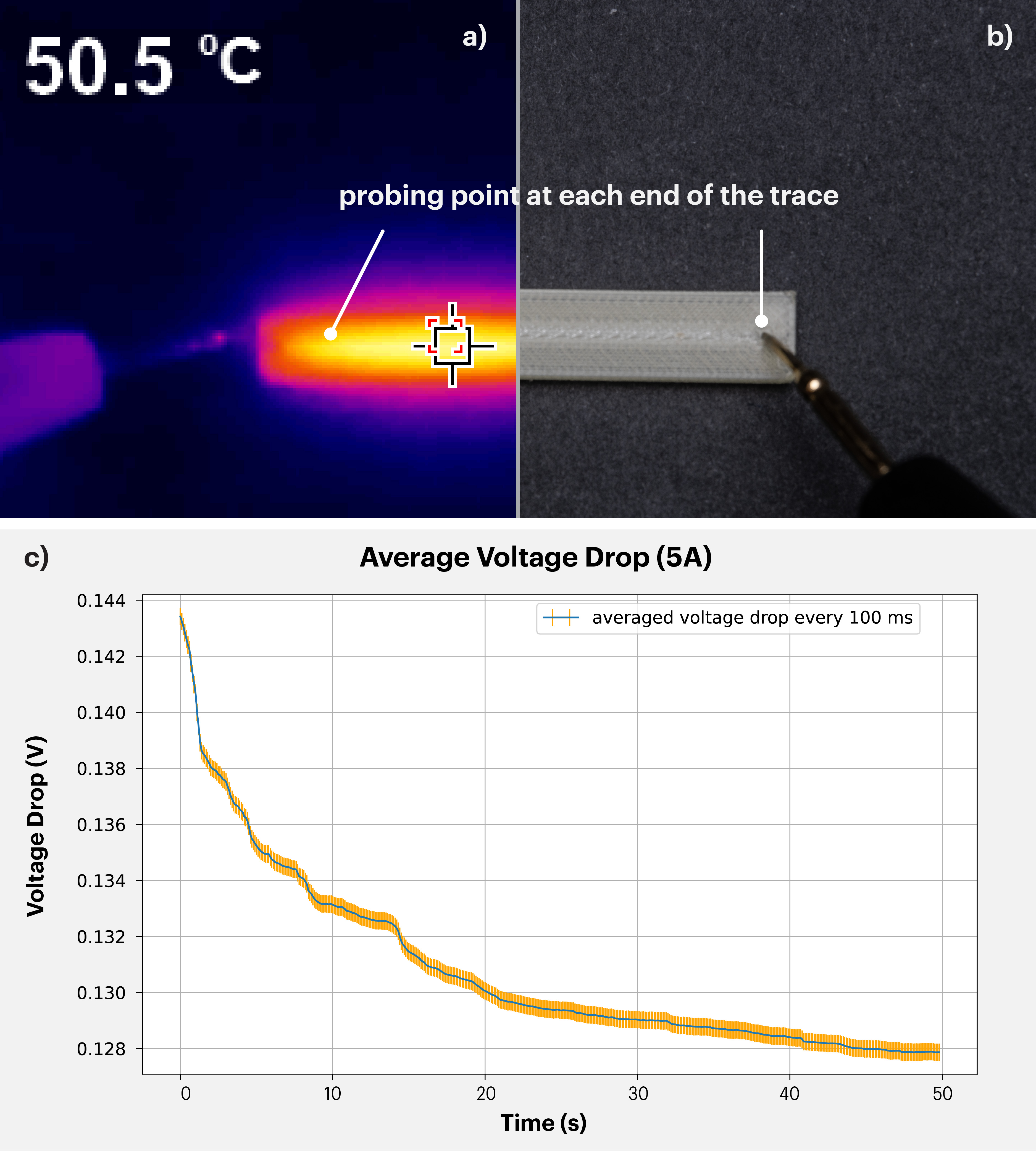}
    \caption{Trace conductivity and current capacity test sample: a) and b) the exterior temperature stabilizes at approximately \SI{50}{\degreeCelsius} under a \SI{5}{\ampere} current load; c) the EGaIn trace voltage drop measured over \SI{50}{\second} under a \SI{5}{\ampere} current.}
    \label{fig:current}
\end{figure}

In addition to conductivity, current-carrying capacity is critical for ensuring circuit reliability and safety under high-current conditions\edited{, which in turn can lead to elevated induction temperatures}. 
Two primary concerns for EGaIn traces operating under such conditions are the potential softening or melting of the PVA substrate and the risk of EGaIn undergoing electromigration due to high current density~\cite{Hillaire2020MarangoniFI}. 
To evaluate the high-current performance of EGaIn traces, we 3D printed \edited{nine} of the thinnest viable traces with a across-section of \SI{0.7}{\milli\meter} $\times$ \SI{0.7}{\milli\meter}. 
Currents of \SI{1}{\ampere}, \SI{3}{\ampere}, and \SI{5}{\ampere} were applied at room temperature \edited{to three samples per condition,} while temperature was monitored using an infrared camera, and both probe openings at the ends were visually inspected for signs of EGaIn overflow. 
\edited{The current capacity test was capped at five minutes, as all samples reached thermal equilibrium during this period, with surface temperatures stabilizing.}
Throughout the test duration, the traces exhibited consistently low voltage drops, with no signs of overheating or EGaIn displacement (Figure \ref{fig:current}b). 
\edited{The average standard deviations in resistance during the current capacity tests were \SI{1.19e-5}{\ohm}, \SI{1.02e-5}{\ohm}, and \SI{1.61e-5}{\ohm} for the \SI{1}{\ampere}, \SI{3}{\ampere}, and \SI{5}{\ampere} conditions, respectively.}
The traces, encapsulated by a \SI{0.3}{\milli\meter}-thick insulating layer on both top and bottom sides, stabilized at a maximum surface temperature of \SI{51}{\celsius} under \SI{5}{\ampere} (Figure \ref{fig:current}a), well below the PVA glass transition temperature range of \SI{85}{\celsius} to \SI{95}{\celsius}.
These results indicate that EGaIn traces can safely support continuous current loads of up to \SI{5}{\ampere} without compromising structural integrity or performance.

\subsection{Waveform Delivery and High-Frequency Performance}

In addition to DC applications, a key consideration in circuit design is the ability to transmit data through high-frequency signals.
To evaluate the high-frequency performance of EGaIn traces and EGaIn-to-component contact points, we fabricated two collinear test traces, each with a \SI{0.7}{\milli\meter} $\times$ \SI{0.7}{\milli\meter} cross-section and a length of \SI{30}{\milli\meter}.
At the midpoint of the traces, we placed a socket for a \SI{0}{\ohm} resistor to connect the two traces.

\begin{figure}[t]
    \centering
    \includegraphics[width=\columnwidth]{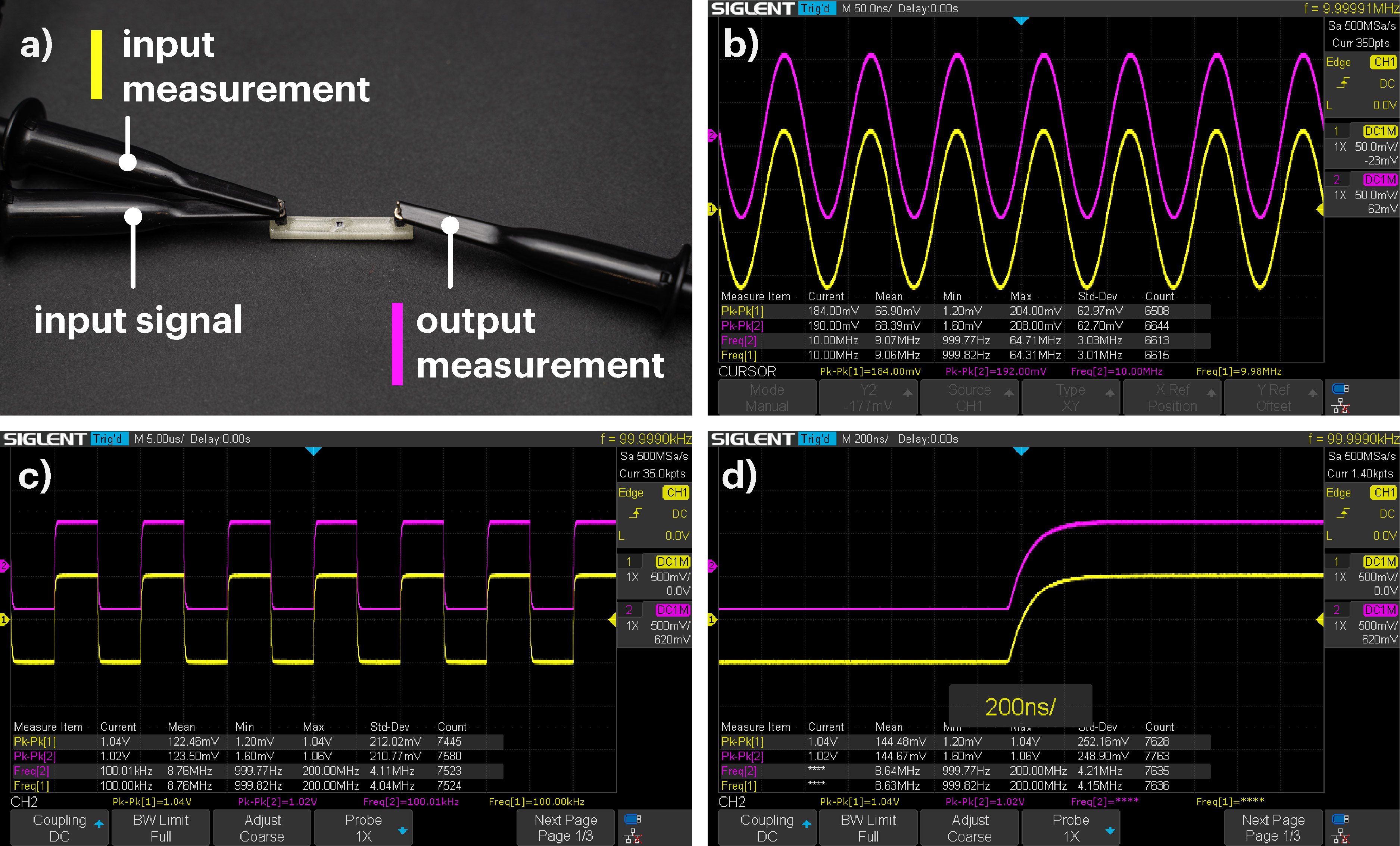}
    \caption{Waveform delivery test sample: a) test setup consisting of two EGaIn traces connected by a \SI{0}{\ohm} resistor at the center; b) input and output sinusoidal waveforms at a \SI{1}{\mega\hertz} frequency; c) input and output quasi-square waveforms at \SI{100}{\kilo\hertz}; d) zoomed-in view of the quasi-square waveform.}
    \label{fig:waveform}
\end{figure}

We supplied both a sinusoidal wave and a quasi-square wave to one end of the trace using a Keysight 33210A Function Generator, with frequencies ranging from \SI{100}{\hertz} to \SI{10}{\mega\hertz} (the maximum supported by the equipment). 
The output waveform was captured from the opposite end of the trace, while the input terminal signal was used as a benchmark, using a SIGLENT SDS 1104X-E oscilloscope. 
The results indicate no noticeable waveform attenuation through the compound trace. 
For example, in the output graphs shown in Figure~\ref{fig:waveform}b–d, both signals—the input and output—demonstrate similar amplitudes and noise levels for both the sine wave and the zoomed-in view of the square wave.
These results show that \dissolvpcb supports high-frequency signal transmission, with EGaIn traces and contacts reliably delivering signals up to \SI{10}{\mega\hertz} without significant signal loss or distortion.

\section{Software} \label{software}
\begin{figure}[b]
    \centering
    \includegraphics[width=\columnwidth]{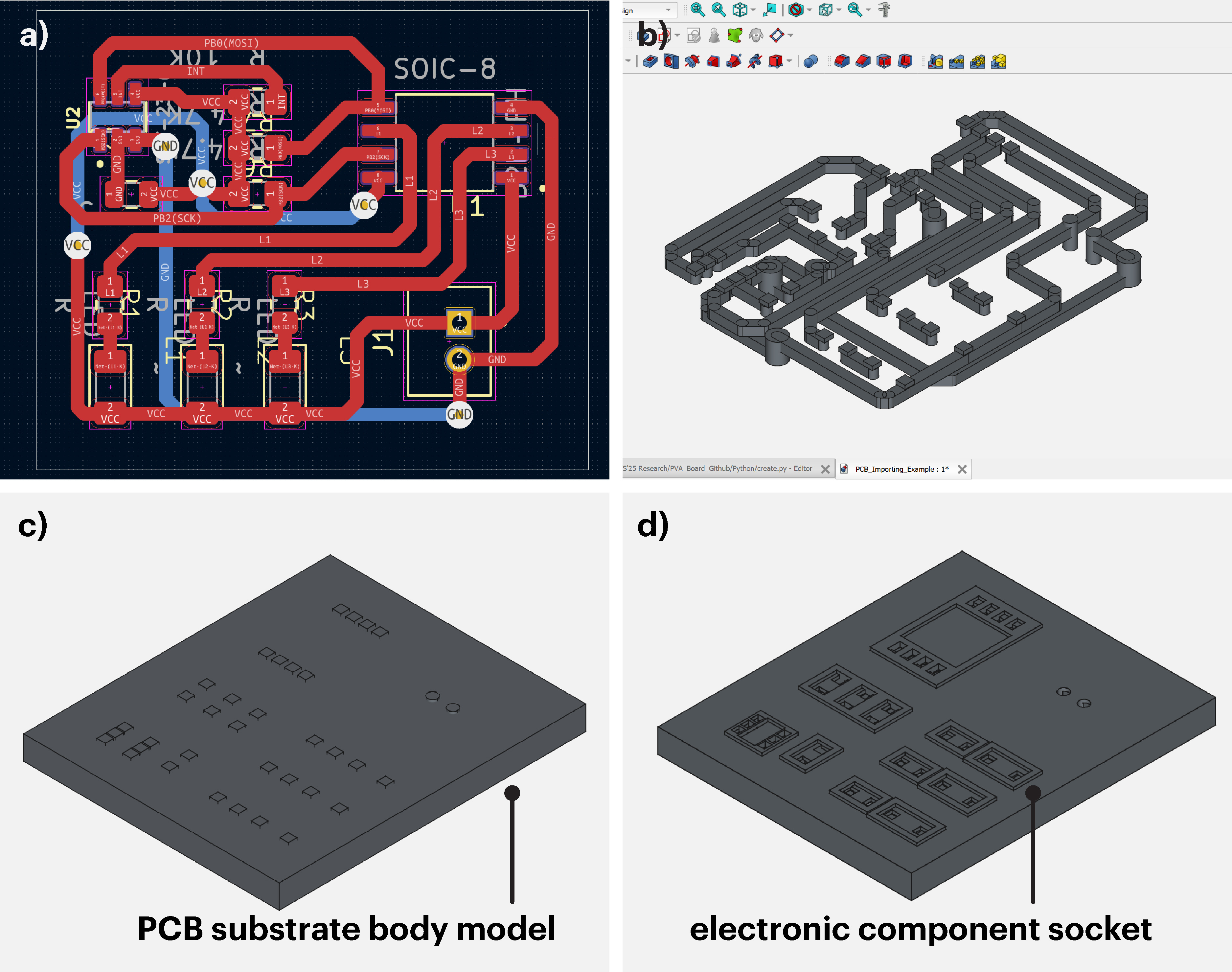}
    \caption{Screenshots of the FreeCAD plugin generating a 3D-printable model for the example circuit: a) PCB design in KiCad; b) 3D model of the traces and pads in FreeCAD; c) 3D model of the substrate body; d) completed 3D substrate model with component sockets.}
    \label{fig:software}

\end{figure}

We now present our open-source FreeCAD plugin, implemented in Python (version 3.13.2) and executed through FreeCAD's built-in macro functionality (Figure~\ref{fig:software}). The plugin automatically converts conventional single- or multi-layer KiCad PCB design files into \dissolvpcb 3D substrate models. To generate a model, users first design the circuit in KiCad and export it as a \texttt{.kicad\_pcb} file. The plugin then parses essential board data from this file and parametrically generates a 3D-printable model tailored to the \dissolvpcb fabrication workflow described in Section~\ref{fab}. The complete plugin is available on GitHub (see Footnote~\ref{fn1}). 

\subsection{KiCad Design Rules and 3D Model Library}\label{drc_model}
Like in traditional PCB workflows, we assume that the main \dissolvpcb layout is created using conventional EDA software such as KiCad. To accommodate the constraints of 3D-printed EGaIn channels and insulation, we provide additional design rule check (DRC) settings to support the \dissolvpcb fabrication process.

Specifically, the minimum trace width is set to \SI{0.7}{\milli\meter}, matching the smallest printable channel width, while both the minimum insulation distance between traces and the minimum conductor-to-board-edge distance are set to \SI{0.15}{\milli\meter}. 
Additionally, we developed a 3D socket model library aligned with the footprint coordinates of common electronic components, including chip packages such as SOIC-8, SOIC-14, and SOT-23, and two-terminal components in 0603, 0805, and 1206 packages.
Users can begin designing PCBs using standard workflows by simply importing the provided DRC rules and custom component library with 3D models. A step-by-step instruction file can be found in the GitHub repository (Footnote \ref{fn1}).

\subsection{Modeling the 3D Substrate} \label{modeling_software}

Our FreeCAD plugin generates a complete 3D model of \dissolvpcb by parsing the PCB design file, generating 3D traces, vias, and pads, constructing the \dissolvpcb base, importing 3D socket models, and performing a global Boolean operation.

\subsubsection{Parsing the PCB file}
During the PCB file parsing step, information regarding the board's physical layout is extracted via keyword matching according to the \texttt{.kicad\_pcb} file structure, which is consistent across all designs. 
The extracted data includes trace segment locations and dimensions, via locations and layer connection information, component footprint positions and orientations, and board polygon dimensions. 
The data is organized into dedicated lists, with each entry stored as a dictionary. 
For example, the list for trace segments contains dictionary entries, each representing a single trace segment with fields for start coordinates, end coordinates, trace width, and board layer.

\subsubsection{Generating traces, vias, and pads}
Using the parsed data on trace segments, vias, and pads, 3D channel and pad models are generated, as shown in Figure~\ref{fig:software}b. 
Compatibility with the \dissolvpcb process is maintained using predefined global parameters such as minimum trace dimensions and layer separation height.

For each trace segment, the plugin calculates the absolute starting coordinates and the displacement along both the X and Y axes based on the length and direction derived from KiCad's start and end coordinates. 
Using this information, it generates a rectangular box with dimensions defined by the user-specified trace width (for both width and height) and the parsed length. 
To ensure seamless transitions between connected segments, a cylinder of matching height and width is added at each end of the box to serve as a joint, eliminating any gaps between adjacent segments. 

Vias are created similarly to trace segment joints, but with a larger diameter of \SI{1.2}{\milli\meter}, as defined in the DRC settings, to ensure optimal EGaIn flow across layers when filling the traces. 
For double-layer PCBs, vias span from the top of the top-layer traces to the bottom of the bottom-layer traces.
For designs with more than two layers, via geometry is generated based on the layer connection data provided in the KiCad file.

Pads are modeled as thin cuboids with a cross-section matching the width of the corresponding trace they connect to and a thickness equal to the minimum Z-direction insulation thickness.
This ensures the Boolean operation creates proper openings to the traces while also preventing pads from being too thin—avoiding insufficient PVA coverage over surface-layer traces.

\subsubsection{Generating the substrate body}
After generating the traces, vias, and pads, the main substrate body is created using the board polygon data, which supports arbitrary shapes composed of line segments and arcs (Figure~\ref{fig:software}c). 
The individual board outline segments, derived from the PCB design file, are sorted into a continuous sequence such that the end coordinate of one segment matches the start coordinate of the next. 
FreeCAD then creates an extrusion from the face formed by the segments.
The total board height is defined by the sum of trace heights and the user-defined insulation thicknesses across all layers.
For example, in the sample circuit, the board height is \SI{2.3}{\milli\meter}, comprising three layers of \SI{0.3}{\milli\meter} Z-direction insulation and two layers of \SI{0.7}{\milli\meter} trace height.

\subsubsection{Placing 3D sockets and final operations}
The 3D socket models from Section~\ref{drc_model} are placed according to the coordinates and orientations specified in their corresponding footprint data (Figure~\ref{fig:software}d). 
This involves traversing the list of collected footprints, identifying the file path for each matching 3D model, importing them into the appropriate locations, and applying the necessary transformations to ensure correct positioning and orientation.

The macro concludes with a global Boolean operation.
The 3D models of traces, vias, and pads are subtracted from the PCB substrate body, while the 3D sockets are joined to it. 

Finally, the completed 3D model can be exported as a mesh file and sliced using the recommended printing parameters described in Section~\ref{fab}.
With the help of the software plugin, the entire process of generating a 3D model from a KiCad PCB design file can be completed in just a few minutes, depending on the complexity of the design and the computer hardware. 
For example, the model for the sample circuit can be generated in under two minutes.

\begin{figure}[!b]
    \centering
    \includegraphics[width=\columnwidth]{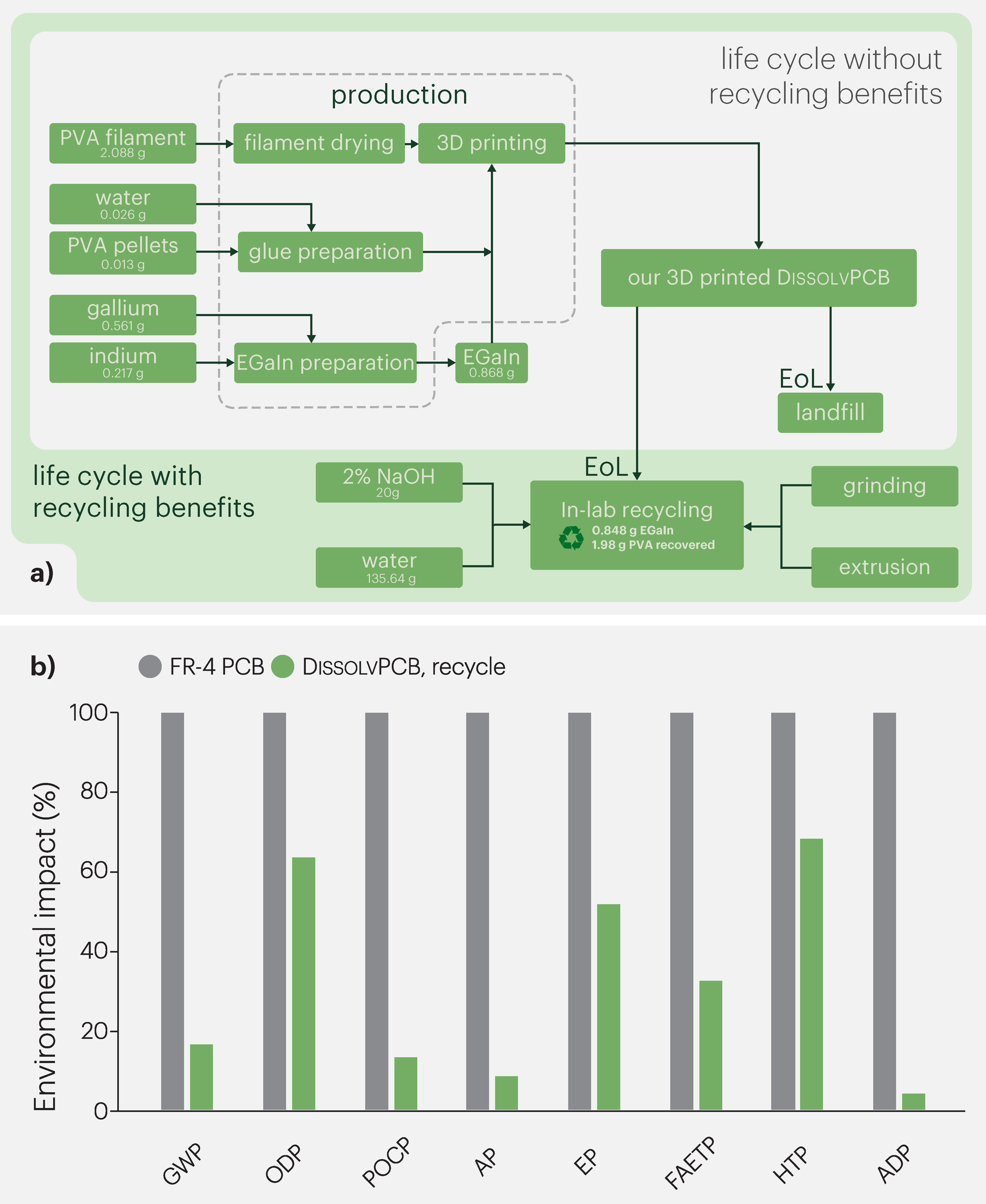}
    \caption{Life cycle assessment: a) life cycle system inventories and boundaries; b) quantitative comparison results of standardized LCIA results between an FR4-based PCB and our \dissolvpcb circuit, including recycling benefits.}
    \label{fig:LCA}

\end{figure}

\section{Assessing the Environmental Impact of \dissolvpcb}\label{lca}

\subsection{Methodology and System Boundary}

To evaluate the environmental sustainability of \dissolvpcb, we conducted a comprehensive cradle-to-grave LCA, focusing on absolute environmental impacts (EIs). 
Our analysis benchmarks two magnetic field detector circuits: one fabricated using \dissolvpcb, and the other using a conventional FR-4 PCB, with identical functionality and dimensions.

The LCA leverages the ecoinvent v3.10 Cutoff database~\cite{ecoinvent3.10}, coupled with the CML v4.8 Life Cycle Impact Assessment (LCIA) methodology~\cite{CMLv4.8}, and is implemented via OpenLCA software~\cite{openLCA}. Due to the unavailability of PVA in the database, we substituted it with PVC (Polyvinyl Chloride). 

The defined system boundary covers all pertinent stages, including material synthesis, manufacturing processes, transportation, and end-of-life management. 
The functional unit was established as ``fabrication of one PCB using the PVA/FR-4 substrate.'' 
The system boundary and procedural flow are illustrated in Figure ~\ref{fig:LCA}a.

\subsection{Material and Energy Inventory}
To fabricate the \dissolvpcb PCBA for the sample circuit described in Section~\ref{overview}, \SI{1.17}{\gram} of PVA filament is used for 3D printing, along with \SI{0.013}{\gram} of PVA pellets and \SI{0.026}{\gram} of water to prepare PVA glue. 
For the conductive traces, \SI{0.651}{\gram} of gallium and \SI{0.217}{\gram} of indium are mixed to form approximately \SI{0.868}{\gram} of EGaIn. 
The manufacturing process includes filament drying (\SI{4.37e-3}{\kilo\watt\hour}), glue preparation (\SI{1.52e-4}{\kilo\watt\hour}), EGaIn synthesis (\SI{5.208e-3}{\kilo\watt\hour}), and 3D printing (\SI{2.7e-2}{\kilo\watt\hour}) to create the final \dissolvpcb board. 

We then carry out the in-lab recycling of the \dissolvpcb PCBA. The process involves immersing the sample in \SI{135.46}{\gram} of water to dissolve the PVA substrate, followed by the application of a 2 wt\% NaOH solution (\SI{20}{\gram}) to recover the liquid metal. This enables the reclamation of approximately \SI{0.848}{\gram} of EGaIn and \SI{1.16}{\gram} of PVA. 
\edited{The reclaimed PVA is then ground and reprocessed into filament using a filament extruder, with the grinding and extrusion steps consuming \SI{2.556e-5}{\kilo\watt\hour} and \SI{4.243e-4}{\kilo\watt\hour}, respectively.} 

In contrast, fabricating the same magnetic field detector board using a conventional approach requires \SI{2.228}{\gram} of FR-4 and \SI{0.628}{\gram} of soldering paste. 
In addition, it consumes \SI{7.0e-4}{\kilo\watt\hour} and \SI{6.68e-3}{\kilo\watt\hour} for soldering and CNC milling, respectively. 

\subsection{Environmental Impact Comparison}
We evaluated eight key environmental indicators to provide an extensive characterization of our approach: Acidification Potential (AP), Eutrophication Potential (EP), Freshwater Aquatic Ecotoxicity Potential (FAETP), Global Warming Potential (GWP), Human Toxic Potential (HTP), Photochemical Ozone Creation Potential (POCP), Abiotic Depletion Potential (ADP, fossil), and Ozone Layer Depletion Potential (ODP). 

Results, depicted in Figure~\ref{fig:LCA}b, show that our \dissolvpcb approach with effective in-lab recycling  significantly outperforms traditional FR-4-based PCBs, exhibiting notably lower environmental impacts across all evaluated metrics.
\edited{
For example, the GWP is reduced to \SI{4.08e-3}{\kilo\gram} CO\textsubscript{2} eq, the ODP to \SI{1.09e-8}{\kilo\gram} CFC-11 eq, the POCP to \SI{3.06e-5}{\kilo\gram} NO\textsubscript{x} eq, and the AP to \SI{2.27e-5}{\kilo\gram} SO\textsubscript{2} eq. 
The EP is measured at \SI{5.02e-6}{\kilo\gram} P eq, the FAETP at \SI{2.27e-4}{\kilo\gram} 1,4-DCB eq, the HTP at \SI{1.49e-2}{\kilo\gram} 1,4-DCB eq, and the ADP at \SI{3.68e-4}{\kilo\gram} oil eq.
All reported values are retained to three significant figures. 
Notably, ODP, HTP, EP, and FAETP are reduced to around 40–70\% of their conventional counterparts, while ADP, AP, and GWP are reduced even further, by approximately one order of magnitude.}

The primary difference in environmental performance between the conventional FR-4 PCB and \dissolvpcb arises from FR-4’s non-recyclability.
Specifically, the process of PCB making and FR-4 production contains intensive use of copper and other resource-heavy materials (e.g., epoxy resins, glass fibers, and Al\textsubscript{2}O\textsubscript{3}) in traditional FR-4 and soldering paste production. The recovering processes of such materials, which include copper extraction, refining, and electroplating, are highly energy-intensive and emit substantial pollutants (e.g., CO\textsubscript{2}, SO\textsubscript{2}, NO\textsubscript{2}, wastewater), thereby significantly contributing to the overall environmental footprint.
These processes are replaced with light-weighted water-based disassembly and separation for easy recovery of primary materials from \dissolvpcb, resulting in significantly lower environmental impact when making the same PCBAs.

\section{examples}
We fabricated three functional circuits to demonstrate the capability of the \dissolvpcb approach.

\begin{figure}[b]
    \centering
    \includegraphics[width=\columnwidth]{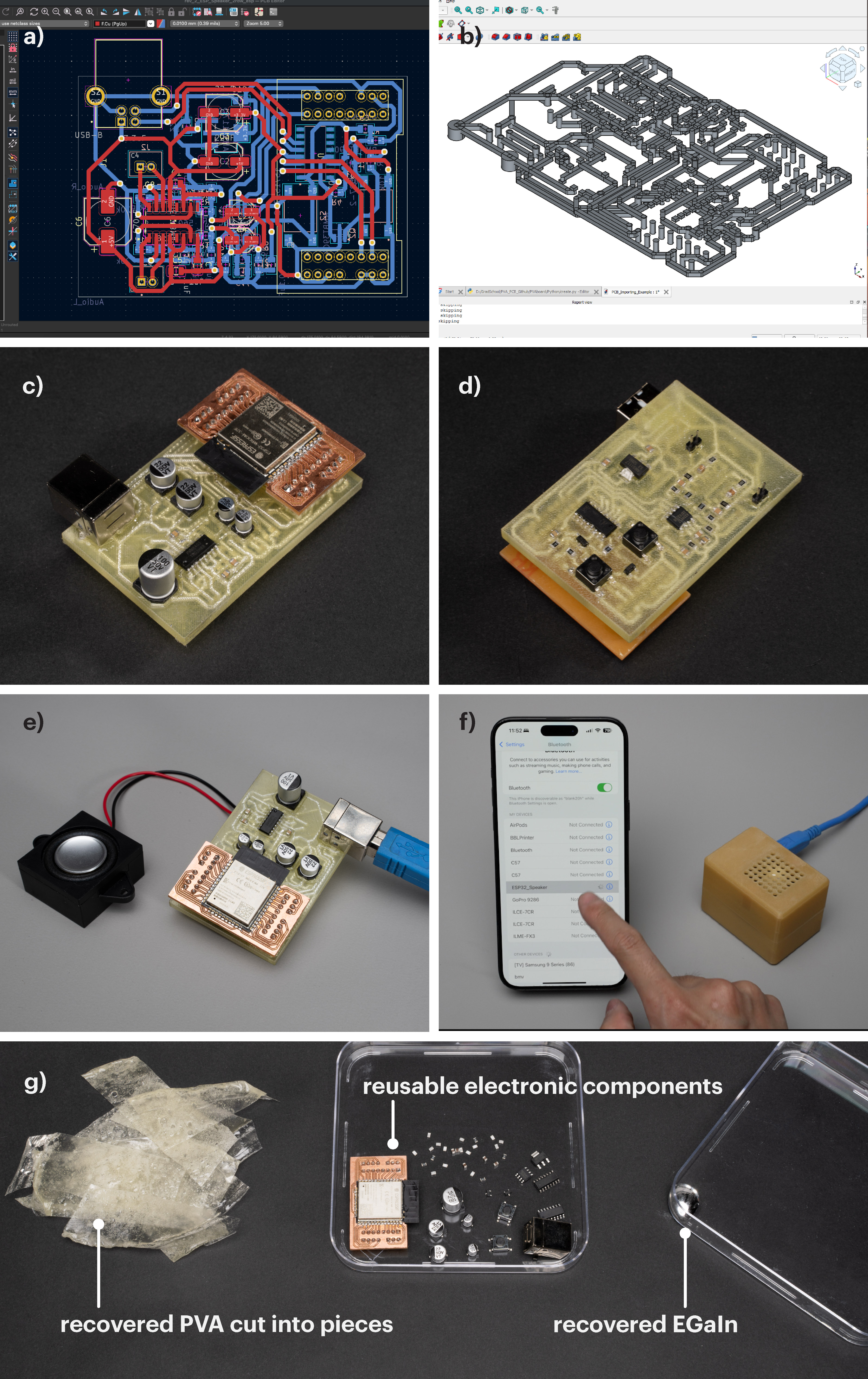}
    \caption{Bluetooth Speaker: a) KiCad design of the speaker circuit; b) software-generated EGaIn channel 3D model (main substrate body hidden); c)–d) front and back sides of the PCB assembly; e)–f) speaker assembly with Bluetooth enabled; g) recovered PVA, EGaIn, and recycled electronic components.}
    \label{fig:speaker}
\end{figure}

\subsection{Recyclable Bluetooth Speaker}\label{speaker}

IoT and smart devices have become increasingly ubiquitous in daily life. 
However, the rapid iteration and development of smart home technologies often result in shortened device lifespans. 
Users frequently replace outdated devices with newer models, contributing to a growing accumulation of e-waste across society. 
The \dissolvpcb approach offers a potential solution by facilitating component recycling through an integrated material recovery process.

In this example, we demonstrate a Bluetooth speaker design featuring a double-layer PCB assembly, as shown in Figure \ref{fig:speaker}.
The design incorporates an ESP32-WROOM-32E (8MB) breakout board, which serves as the central computing and communication controller (Figure \ref{fig:speaker}a).
This is complemented by SMD peripherals including a UART interface IC, audio amplifier, digital-to-analog converter (DAC), voltage regulator, and 31 two-terminal SMD components (resistors and capacitors) of various sizes and values. 
These elements are arranged to optimize signal processing, power management, and subfunction integration. 
The 3D-printed substrate is automatically generated by the software plugin (Figure \ref{fig:speaker}b).
The fabricated Bluetooth speaker can be paired with a smartphone to broadcast music (Figure \ref{fig:speaker}c–f).

Despite the functional complexity of this PCB assembly, its recycling process follows the same streamlined procedure outlined in Section~\ref{recycle}. 
Through this method, we recovered 99.4\% of the \SI{10.13}{\gram} PVA and 98.6\% of the \SI{3.64}{\gram} liquid metal used in the speaker PCBA (Figure \ref{fig:speaker}g). 
This result underscores the feasibility of manufacturing future personal devices using PVA and LM composites, which can later be efficiently collected, recycled, and repurposed with minimal effort. 
Such an approach not only reduces e-waste but also advances closed-loop sustainable manufacturing by reclaiming raw materials for future production cycles.

\begin{figure}[t]
    \centering
    \includegraphics[width = \columnwidth]{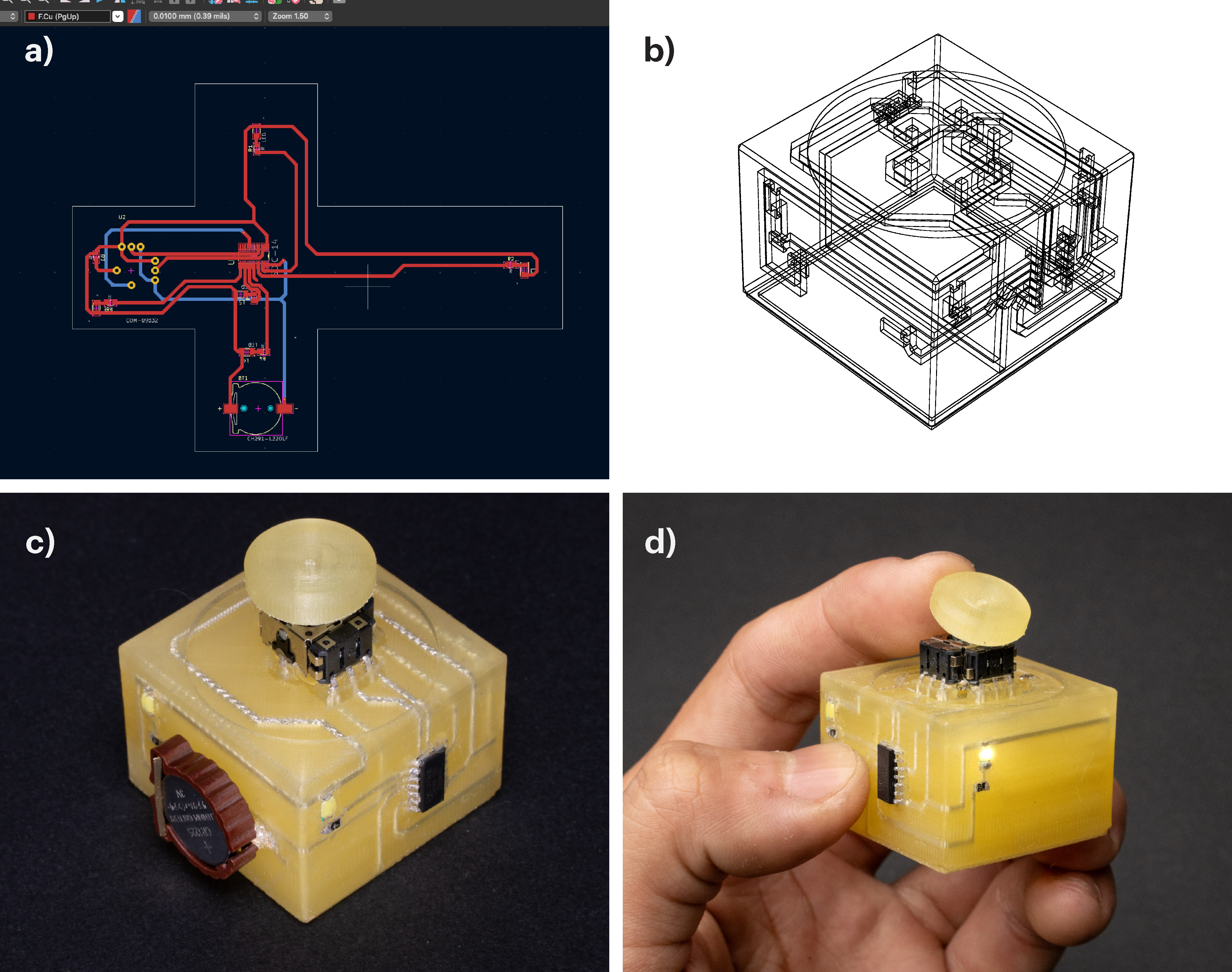}
    \caption{Electronic Fidget: a) the KiCad PCB design used as a reference; b) the wireframe view of the fidget's 3D circuit model; c) the fidget assembly; d) interacting with the electronic fidget, with the joystick activating an LED on the corresponding side.}
    \label{fig:fidget}
\end{figure}

\subsection{Electronic Fidget}\label{fidget}
While the primary contribution of \dissolvpcb lies in enabling the creation of fully recyclable PCBAs, the use of 3D printing for substrate fabrication offers additional benefits. 
For example, conductive traces can be routed not only in two dimensions but also extended into three dimensions, enabling customized, electronics-integrated structures that conventional flat FR-4 PCBs cannot support.

To demonstrate this capability, we present a 3D-printed fidget gadget that responds to mechanical inputs via a joystick while providing dynamic feedback through programmable LED lighting, similar to a ``wack-a-mole'' game.
The gadget features a cubic form factor, with the joystick mounted on the top surfaces and one SMD LED embedded on each of the four vertical sides (Figure~\ref{fig:fidget}c–d). 
The embedded circuitry employs an ATtiny microcontroller supported by nine peripheral components to manage input detection and output behavior. 
The 3D-printed substrate, fabricated as a single piece, includes integrated component sockets and conductive trace channels across all five surfaces. 

The circuit design process begins with a conventional flat layout (Figure~\ref{fig:fidget}a). 
We then manually map the traces and components onto the surfaces of the cubic structure using a CAD modeling tool (Figure~\ref{fig:fidget}b), as our current plugin does not support direct 3D routing conversion. 
However, we note that recent computational tools such as ModElec~\cite{ModElec} may support direct 3D circuit layout and could be integrated into our design pipeline in future iterations.

Following the recycling protocol described in Section \ref{recycle}, we recovered 99.1\% of the \SI{29.92}{\gram} PVA used and 97.7\% of the \SI{2.09}{\gram} of liquid metal used in the fidget device.

\begin{figure}[b]
    \centering
    \includegraphics[width = \columnwidth]{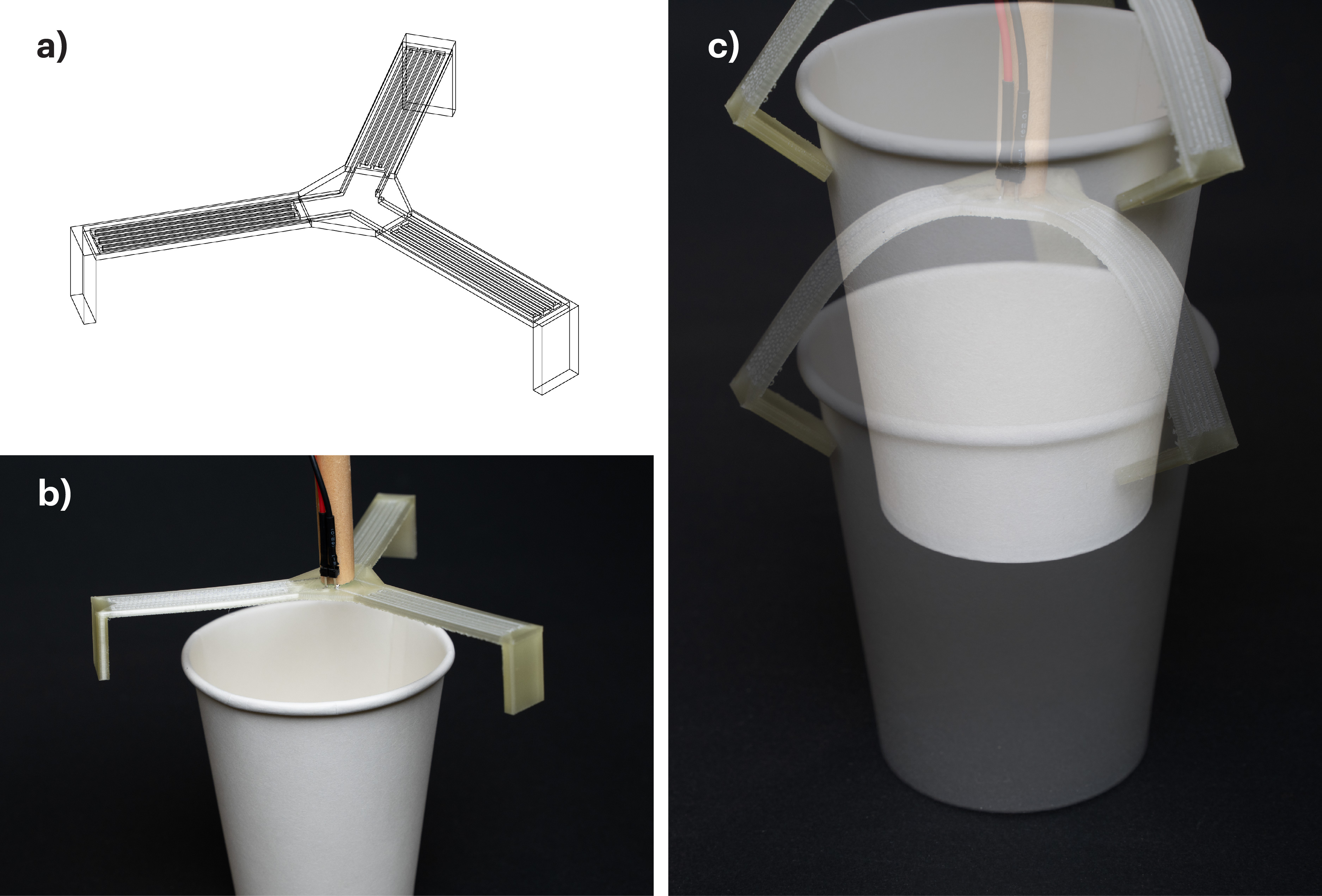}
    \caption{Three-finger gripper: a)  gripper design with extended conductive traces for Joule heating activation; b) initial state prior to activation; c) deformation triggered by Joule heating, gripping a cup.}
    \label{fig:gripper}
\end{figure} 

\subsection{4D Printed Three-Finger Gripper}\label{gripper}
Our final example demonstrates the additional advantages of integrating liquid metal conductors in circuit design. Since changes in a circuit's shape do not compromise the continuity or conductivity of EGaIn-embedded traces, \dissolvpcb can also be used to prototype shape-changing artifacts, a process known as 4D printing~\cite{MOMENI201742}. 
In contrast to many previous examples where morphing behavior relies on external heat sources (e.g., hot water baths~\cite{4DMesh, Geodesy}), embedded EGaIn enables direct Joule heating when current is applied. 

To demonstrate this capability, we fabricated a three-finger, shape-changing gripper that was 3D printed with internal channels filled with EGaIn (Figure~\ref{fig:gripper}). 
When energized, the traces generate heat, inducing controlled deformation of the surrounding PVA polymer matrix. 
This example, while simple, highlights the potential for applications such as flat-pack items that self-assemble upon electrical activation, adaptive wearable devices for rehabilitation, and reconfigurable architectural components, all deployable in compact form and activated on demand. When no longer needed, the entire device can be dissolved and recycled in water.

As with previous examples, we recycled the finger gripper and recovered 98.7\% of the \SI{6.68}{\gram} PVA and 99.0\% of the \SI{6.12}{\gram} liquid metal used.

\section{Discussion}
\dissolvpcb is the first pipeline that enables the design and fabrication of reliable and fully recyclable PCBAs that are functionally comparable to traditional FR-4 PCBs. 
However, like all novel systems and techniques, employing \dissolvpcb comes with its own trade-offs and limitations. 
We now discuss some of these considerations and outline potential future directions.

\subsection{Size of a PCBA}
Our lab experiments and prototypes have demonstrated that \dissolvpcb can produce sustainable PCBAs for a range of applications, including logic circuits and those requiring high current. 
Compared to FR-4 PCBs, one major limitation of \dissolvpcb is the potential increase in board footprint, primarily due to constraints in trace channel dimensions. 
For example, the minimum printable trace channel width in \dissolvpcb is \SI{0.7}{\milli\meter}, whereas in an FR-4 PCB with \SI{1}{\oz\per\sqft} copper thickness, the recommended trace width for a \SI{0.5}{\ampere} application is \SI{15}{\mil}, which is approximately half the width required by \dissolvpcb. 
Therefore, for applications with strict size constraints, \dissolvpcb may not be ideal.

A potential solution is to take advantage of 3D trace routing, allowing a traditional FR-4 circuit layout to be restructured into a \dissolvpcb design with a similar footprint by increasing only the vertical dimension. 
With our custom DRC files, KiCad's built-in Design Rule Checker tool can verify whether a PCB design complies with the recommended printing parameters.

\subsection{Applications Exposed to Water}
Another limitation of \dissolvpcb is that the fabricated PCBA cannot be exposed to water, which may restrict its use in applications such as wearables that are subject to various environmental conditions. 
One potential solution is to employ enclosure designs that incorporate gaskets or compression seals. 
Modern consumer electronic devices often achieve IPX6 ratings~\cite{iec60529} using similar techniques, even though the PCBAs inside these devices are not inherently waterproof.

Another related concern is the lifespan of the bare PVA-based PCBA, which is often required during prototyping. 
While we do not yet have systematic data to report in this regard, the sample circuits shown in Figure~\ref{fig:benchmark} have been functioning for over 60 days in a typical indoor environment, where the humidity fluctuated between 19\% and 65\% due to weather changes. 
All fabricated circuits continued to operate properly without visible deformation.

\subsection{Future Technical Directions}
\edited{While \dissolvpcb\ enables fully recyclable PCBAs with diverse components and custom form factors, several future improvements could further enhance its capabilities.
}

\subsubsection{Component compatibility.} \edited{In section \ref{socket}, we classify electronic components compatible with \dissolvpcb based on their footprint packages. 
However, we acknowledge that component compatibility extends beyond physical fit and warrants a more systematic investigation. 
For example, while plastic-encapsulated ICs and ceramic two-terminal components integrate well with \dissolvpcb, components with high moisture sensitivity may require additional protective measures. 
Expanding the study to include a wider variety of electronic component types could yield insights for advancing both \dissolvpcb and general solvent-based electronics recycling technologies.}

\subsubsection{Software to support 3D routing.} \edited{The current software implementation focuses on parsing conventional single- and multi-layer PCB designs with standard form factors. Looking ahead, we see great potential in expanding it toward an EDA framework that supports three-dimensional design and routing. Prior work has explored 3D routing for modeling purposes~\cite{ModElec, A_series_of_tubes}, and we envision that integrating similar techniques into a schematic-based 3D routing EDA could greatly enhance the accessibility and adoption of \dissolvpcb.}

\subsubsection{Debugging and repairability.} \edited{The current workflow does not explicitly support debugging after fabrication. However, since \dissolvpcb\ accommodates THT components, it is feasible to introduce test pads in the future, either as THT pogo-pin receivers for automated testing or as standard THT pads for manual probing. 
}

\edited{Future work may also investigate the repairability of PVA-based circuits. For example, by tuning the sensitivity of the PVA glue~\cite{chen2021quantitative}, it may be possible to exploit differences in water absorption rates between the glue and the printed substrate to enable localized removal and replacement of electronic components.
}

\subsubsection{Toward full automation.}
The \dissolvpcb\ technique is designed for use in everyday makerspaces and can be adopted without significant additional hardware investment. To support this goal, the current workflow involves the manual injection of EGaIn and sealing of components with PVA glue. While the process is straightforward, it involves hands-on tasks that may require practice to perform effectively.
Alternatively, we envision that these manual efforts could be reduced by equipping the 3D printing platform with an additional syringe extruder. 
Previous research has shown that such hardware modifications are low-cost and compatible with commercial FDM 3D printers~\cite{Magnetophoretic}. 
With similar hardware, future work may enable full automation of the fabrication process.

\subsection{An Alternative Recycling Ecosystem?}

As \dissolvpcb has the potential to democratize the creation and use of fully recyclable PCBAs, it opens the door to an alternative e-waste recycling paradigm. 
Instead of transporting dispersed e-waste to a few centralized, high-energy-cost recycling facilities, which typically recover only certain raw metals, \dissolvpcb presents a model where much of the e-waste can be processed locally, with lower energy consumption, significantly higher recovery yield, and the preservation of all functional electronic components. 
These components, along with EGaIn and PVA scraps, can then be directly returned to manufacturing pipelines.

Inevitably, this alternative recycling paradigm raises a number of new questions. 
For example, permanent soldering practices and densely packed trace layouts are sometimes used to protect the intellectual property of a design, such as by deterring reverse engineering. 
Increased recyclability may challenge these intentions. 
Additionally, how to motivate makers and consumers to take extra action in recycling e-waste remains an open question. 
Addressing these issues requires research in environmental science and political science, along with broader interdisciplinary discussions that extend beyond the scope of a single paper. 
Nevertheless, we view \dissolvpcb as a meaningful step toward a more sustainable future in electronics manufacturing.

\section{Conclusion}
In this paper, we presented \dissolvpcb, a fully recyclable PCBA approach that combines PVA-based FDM 3D printing with EGaIn liquid metal circuitry. This method enables easy disassembly and component recovery through simple water immersion. We detailed the fabrication and recycling workflow, evaluated its performance and environmental impact, and introduced an open-source plugin that converts EDA designs into printable models. We demonstrated three functional prototypes made using the \dissolvpcb approach and discussed its technical limitations. We concluded with potential directions for future research.

\begin{acks}
We thank Sandbox, the Jagdeep Singh Family Makerspace, Terrapin Works, and BioWorkshop for providing access to tools during the development and documentation of this project. This research was supported in part by the National Science Foundation under award numbers CNS-2430327 and CNS-2324861. We also gratefully acknowledge support from the Alfred P. Sloan Foundation, VMware, and Google. 
Any opinions, findings, conclusions, or recommendations expressed in this material are those of the authors and do not necessarily reflect the views of the National Science Foundation or other supporters. A large language model service was used solely for proofreading purposes.
\end{acks}

\bibliographystyle{ACM-Reference-Format}
\bibliography{sample-base}


\begin{thebibliography}{76}


\ifx \showCODEN    \undefined \def \showCODEN     #1{\unskip}     \fi
\ifx \showDOI      \undefined \def \showDOI       #1{#1}\fi
\ifx \showISBNx    \undefined \def \showISBNx     #1{\unskip}     \fi
\ifx \showISBNxiii \undefined \def \showISBNxiii  #1{\unskip}     \fi
\ifx \showISSN     \undefined \def \showISSN      #1{\unskip}     \fi
\ifx \showLCCN     \undefined \def \showLCCN      #1{\unskip}     \fi
\ifx \shownote     \undefined \def \shownote      #1{#1}          \fi
\ifx \showarticletitle \undefined \def \showarticletitle #1{#1}   \fi
\ifx \showURL      \undefined \def \showURL       {\relax}        \fi
\providecommand\bibfield[2]{#2}
\providecommand\bibinfo[2]{#2}
\providecommand\natexlab[1]{#1}
\providecommand\showeprint[2][]{arXiv:#2}

\bibitem[\protect\citeauthoryear{Arroyos, Viitaniemi, Keehn, Oruganti, Saunders, Strauss, Iyer, and Nguyen}{Arroyos et~al\mbox{.}}{2022}]%
        {A_Tale_of_Two_Mice}
\bibfield{author}{\bibinfo{person}{Vicente Arroyos}, \bibinfo{person}{Maria L~K Viitaniemi}, \bibinfo{person}{Nicholas Keehn}, \bibinfo{person}{Vaidehi Oruganti}, \bibinfo{person}{Winston Saunders}, \bibinfo{person}{Karin Strauss}, \bibinfo{person}{Vikram Iyer}, {and} \bibinfo{person}{Bichlien~H Nguyen}.} \bibinfo{year}{2022}\natexlab{}.
\newblock \showarticletitle{A Tale of Two Mice: Sustainable Electronics Design and Prototyping}. In \bibinfo{booktitle}{\emph{Extended Abstracts of the 2022 CHI Conference on Human Factors in Computing Systems}} (New Orleans, LA, USA) \emph{(\bibinfo{series}{CHI EA '22})}. \bibinfo{publisher}{Association for Computing Machinery}, \bibinfo{address}{New York, NY, USA}, Article \bibinfo{articleno}{263}, \bibinfo{numpages}{10}~pages.
\newblock
\showISBNx{9781450391566}
\urldef\tempurl%
\url{https://doi.org/10.1145/3491101.3519823}
\showDOI{\tempurl}


\bibitem[\protect\citeauthoryear{Baldé, Kuehr, Yamamoto, McDonald, D’Angelo, Althaf, Bel, Deubzer, Fernandez-Cubillo, Forti, Gray, Herat, Honda, Iattoni, Khetriwal, Luda~di Cortemiglia, Lobuntsova, Nnorom, Pralat, and Wagner}{Baldé et~al\mbox{.}}{2024}]%
        {balde2024global}
\bibfield{author}{\bibinfo{person}{Cornelis~P. Baldé}, \bibinfo{person}{Ruediger Kuehr}, \bibinfo{person}{Tales Yamamoto}, \bibinfo{person}{Rosie McDonald}, \bibinfo{person}{Elena D’Angelo}, \bibinfo{person}{Shahana Althaf}, \bibinfo{person}{Garam Bel}, \bibinfo{person}{Otmar Deubzer}, \bibinfo{person}{Elena Fernandez-Cubillo}, \bibinfo{person}{Vanessa Forti}, \bibinfo{person}{Vanessa Gray}, \bibinfo{person}{Sunil Herat}, \bibinfo{person}{Shunichi Honda}, \bibinfo{person}{Giulia Iattoni}, \bibinfo{person}{Deepali~S. Khetriwal}, \bibinfo{person}{Vittoria Luda~di Cortemiglia}, \bibinfo{person}{Yuliya Lobuntsova}, \bibinfo{person}{Innocent Nnorom}, \bibinfo{person}{Noémie Pralat}, {and} \bibinfo{person}{Michelle Wagner}.} \bibinfo{year}{2024}\natexlab{}.
\newblock \bibinfo{title}{Global E-waste Monitor 2024}.
\newblock
\newblock


\bibitem[\protect\citeauthoryear{Bell, Al~Naimi, McQuaid, and Alistar}{Bell et~al\mbox{.}}{2022}]%
        {Alganyl}
\bibfield{author}{\bibinfo{person}{Fiona Bell}, \bibinfo{person}{Latifa Al~Naimi}, \bibinfo{person}{Ella McQuaid}, {and} \bibinfo{person}{Mirela Alistar}.} \bibinfo{year}{2022}\natexlab{}.
\newblock \showarticletitle{Designing with Alganyl}. In \bibinfo{booktitle}{\emph{Sixteenth International Conference on Tangible, Embedded, and Embodied Interaction}} (Daejeon, Republic of Korea) \emph{(\bibinfo{series}{TEI '22})}. \bibinfo{publisher}{Association for Computing Machinery}, \bibinfo{address}{New York, NY, USA}, Article \bibinfo{articleno}{2}, \bibinfo{numpages}{14}~pages.
\newblock
\showISBNx{9781450391474}
\urldef\tempurl%
\url{https://doi.org/10.1145/3490149.3501308}
\showDOI{\tempurl}


\bibitem[\protect\citeauthoryear{Bell, Chow, Choi, and Alistar}{Bell et~al\mbox{.}}{2023}]%
        {SCOBY}
\bibfield{author}{\bibinfo{person}{Fiona Bell}, \bibinfo{person}{Derrek Chow}, \bibinfo{person}{Hyelin Choi}, {and} \bibinfo{person}{Mirela Alistar}.} \bibinfo{year}{2023}\natexlab{}.
\newblock \showarticletitle{SCOBY BREASTPLATE: SLOWLY GROWING A MICROBIAL INTERFACE}. In \bibinfo{booktitle}{\emph{Proceedings of the Seventeenth International Conference on Tangible, Embedded, and Embodied Interaction}} (Warsaw, Poland) \emph{(\bibinfo{series}{TEI '23})}. \bibinfo{publisher}{Association for Computing Machinery}, \bibinfo{address}{New York, NY, USA}, Article \bibinfo{articleno}{34}, \bibinfo{numpages}{15}~pages.
\newblock
\showISBNx{9781450399777}
\urldef\tempurl%
\url{https://doi.org/10.1145/3569009.3572805}
\showDOI{\tempurl}


\bibitem[\protect\citeauthoryear{Bharath, Madhu, Gowda, Verma, Sanjay, and Siengchin}{Bharath et~al\mbox{.}}{2020}]%
        {bharath2020novel}
\bibfield{author}{\bibinfo{person}{Kurki~N. Bharath}, \bibinfo{person}{Puttegowda Madhu}, \bibinfo{person}{Thyavihalli G.~Y. Gowda}, \bibinfo{person}{Akarsh Verma}, \bibinfo{person}{Mavinkere~R. Sanjay}, {and} \bibinfo{person}{Suchart Siengchin}.} \bibinfo{year}{2020}\natexlab{}.
\newblock \showarticletitle{A novel approach for development of printed circuit board from biofiber based composites}.
\newblock \bibinfo{journal}{\emph{Polymer Composites}} \bibinfo{volume}{41}, \bibinfo{number}{11} (\bibinfo{year}{2020}), \bibinfo{pages}{4550--4558}.
\newblock
\urldef\tempurl%
\url{https://doi.org/10.1002/pc.25732}
\showDOI{\tempurl}


\bibitem[\protect\citeauthoryear{Biswal, Hong, Zhang, Zheng, Gupta, Nepal, Iyer, and Vashisth}{Biswal et~al\mbox{.}}{2025}]%
        {Biswal2025}
\bibfield{author}{\bibinfo{person}{Agni~K. Biswal}, \bibinfo{person}{Peter Hong}, \bibinfo{person}{Zhihan Zhang}, \bibinfo{person}{Yiwen Zheng}, \bibinfo{person}{Surabhit Gupta}, \bibinfo{person}{Dhriti Nepal}, \bibinfo{person}{Vikram Iyer}, {and} \bibinfo{person}{Aniruddh Vashisth}.} \bibinfo{year}{2025}\natexlab{}.
\newblock \showarticletitle{Flexible and Stretchable Vitrimers for Sustainable Electronics}.
\newblock \bibinfo{journal}{\emph{ACS Applied Materials \& Interfaces}} \bibinfo{volume}{17}, \bibinfo{number}{6} (\bibinfo{year}{2025}), \bibinfo{pages}{9736--9747}.
\newblock
\showISSN{1944-8244}
\urldef\tempurl%
\url{https://doi.org/10.1021/acsami.4c16995}
\showDOI{\tempurl}


\bibitem[\protect\citeauthoryear{Blevis}{Blevis}{2007}]%
        {SID}
\bibfield{author}{\bibinfo{person}{Eli Blevis}.} \bibinfo{year}{2007}\natexlab{}.
\newblock \showarticletitle{Sustainable Interaction Design: Invention \& Disposal, Renewal \& Reuse}. In \bibinfo{booktitle}{\emph{Proceedings of the SIGCHI Conference on Human Factors in Computing Systems}} (San Jose, California, USA) \emph{(\bibinfo{series}{CHI '07})}. \bibinfo{publisher}{Association for Computing Machinery}, \bibinfo{address}{New York, NY, USA}, \bibinfo{pages}{503–512}.
\newblock
\showISBNx{9781595935939}
\urldef\tempurl%
\url{https://doi.org/10.1145/1240624.1240705}
\showDOI{\tempurl}


\bibitem[\protect\citeauthoryear{Buechley, Ta, and Johnson}{Buechley et~al\mbox{.}}{2023}]%
        {Printable_Play-Dough}
\bibfield{author}{\bibinfo{person}{Leah Buechley}, \bibinfo{person}{Ruby Ta}, {and} \bibinfo{person}{Alyssa Johnson}.} \bibinfo{year}{2023}\natexlab{}.
\newblock \showarticletitle{3D Printable Play-Dough}. In \bibinfo{booktitle}{\emph{Extended Abstracts of the 2023 CHI Conference on Human Factors in Computing Systems}} \emph{(\bibinfo{series}{CHI EA '23})}. \bibinfo{publisher}{Association for Computing Machinery}, \bibinfo{address}{New York, NY, USA}, Article \bibinfo{articleno}{428}, \bibinfo{numpages}{4}~pages.
\newblock
\showISBNx{9781450394222}
\urldef\tempurl%
\url{https://doi.org/10.1145/3544549.3583927}
\showDOI{\tempurl}


\bibitem[\protect\citeauthoryear{Chen, Yang, Huang, Ge, Yao, Tang, Ren, Ren, and Ma}{Chen et~al\mbox{.}}{2021}]%
        {chen2021quantitative}
\bibfield{author}{\bibinfo{person}{Siqi Chen}, \bibinfo{person}{Hao Yang}, \bibinfo{person}{Kui Huang}, \bibinfo{person}{Xiaolong Ge}, \bibinfo{person}{Hanpeng Yao}, \bibinfo{person}{Junxiang Tang}, \bibinfo{person}{Junxue Ren}, \bibinfo{person}{Shixue Ren}, {and} \bibinfo{person}{Yanli Ma}.} \bibinfo{year}{2021}\natexlab{}.
\newblock \showarticletitle{Quantitative study on solubility parameters and related thermodynamic parameters of PVA with different alcoholysis degrees}.
\newblock \bibinfo{journal}{\emph{Polymers}} \bibinfo{volume}{13}, \bibinfo{number}{21} (\bibinfo{year}{2021}), \bibinfo{pages}{3778}.
\newblock


\bibitem[\protect\citeauthoryear{Cheng, Narumi, Do, Zhang, Ta, Sasatani, Markvicka, Kawahara, Yao, Abowd, and Oh}{Cheng et~al\mbox{.}}{2020}]%
        {Silver_Tape}
\bibfield{author}{\bibinfo{person}{Tingyu Cheng}, \bibinfo{person}{Koya Narumi}, \bibinfo{person}{Youngwook Do}, \bibinfo{person}{Yang Zhang}, \bibinfo{person}{Tung~D. Ta}, \bibinfo{person}{Takuya Sasatani}, \bibinfo{person}{Eric Markvicka}, \bibinfo{person}{Yoshihiro Kawahara}, \bibinfo{person}{Lining Yao}, \bibinfo{person}{Gregory~D. Abowd}, {and} \bibinfo{person}{HyunJoo Oh}.} \bibinfo{year}{2020}\natexlab{}.
\newblock \showarticletitle{Silver Tape: Inkjet-Printed Circuits Peeled-and-Transferred on Versatile Substrates}.
\newblock \bibinfo{journal}{\emph{Proc. ACM Interact. Mob. Wearable Ubiquitous Technol.}} \bibinfo{volume}{4}, \bibinfo{number}{1}, Article \bibinfo{articleno}{6} (\bibinfo{date}{mar} \bibinfo{year}{2020}), \bibinfo{numpages}{17}~pages.
\newblock
\urldef\tempurl%
\url{https://doi.org/10.1145/3381013}
\showDOI{\tempurl}


\bibitem[\protect\citeauthoryear{Cheng, Tabb, Park, Gallo, Maheshwari, Abowd, Oh, and Danielescu}{Cheng et~al\mbox{.}}{2023}]%
        {cheng2023functional}
\bibfield{author}{\bibinfo{person}{Tingyu Cheng}, \bibinfo{person}{Taylor Tabb}, \bibinfo{person}{Jung~Wook Park}, \bibinfo{person}{Eric~M Gallo}, \bibinfo{person}{Aditi Maheshwari}, \bibinfo{person}{Gregory~D. Abowd}, \bibinfo{person}{Hyunjoo Oh}, {and} \bibinfo{person}{Andreea Danielescu}.} \bibinfo{year}{2023}\natexlab{}.
\newblock \showarticletitle{Functional Destruction: Utilizing Sustainable Materials’ Physical Transiency for Electronics Applications}. In \bibinfo{booktitle}{\emph{Proceedings of the 2023 CHI Conference on Human Factors in Computing Systems}} (Hamburg, Germany) \emph{(\bibinfo{series}{CHI '23})}. \bibinfo{publisher}{Association for Computing Machinery}, \bibinfo{address}{New York, NY, USA}, Article \bibinfo{articleno}{366}, \bibinfo{numpages}{16}~pages.
\newblock
\showISBNx{9781450394215}
\urldef\tempurl%
\url{https://doi.org/10.1145/3544548.3580811}
\showDOI{\tempurl}


\bibitem[\protect\citeauthoryear{Cheng, Zhang, Huang, Gao, Sun, Abowd, Oh, and Hester}{Cheng et~al\mbox{.}}{2024}]%
        {cheng2024recy}
\bibfield{author}{\bibinfo{person}{Tingyu Cheng}, \bibinfo{person}{Zhihan Zhang}, \bibinfo{person}{Han Huang}, \bibinfo{person}{Yingting Gao}, \bibinfo{person}{Wei Sun}, \bibinfo{person}{Gregory~D Abowd}, \bibinfo{person}{HyunJoo Oh}, {and} \bibinfo{person}{Josiah Hester}.} \bibinfo{year}{2024}\natexlab{}.
\newblock \showarticletitle{Recy-ctronics: Designing Fully Recyclable Electronics With Varied Form Factors}.
\newblock \bibinfo{journal}{\emph{arXiv preprint arXiv:2406.09611}} (\bibinfo{year}{2024}).
\newblock


\bibitem[\protect\citeauthoryear{Cui and Forssberg}{Cui and Forssberg}{2003}]%
        {cui2003mechanical}
\bibfield{author}{\bibinfo{person}{Jirang Cui} {and} \bibinfo{person}{Eric Forssberg}.} \bibinfo{year}{2003}\natexlab{}.
\newblock \showarticletitle{Mechanical recycling of waste electric and electronic equipment: a review}.
\newblock \bibinfo{journal}{\emph{Journal of hazardous materials}} \bibinfo{volume}{99}, \bibinfo{number}{3} (\bibinfo{year}{2003}), \bibinfo{pages}{243--263}.
\newblock


\bibitem[\protect\citeauthoryear{Deshpande, Takahashi, and Kim}{Deshpande et~al\mbox{.}}{2024}]%
        {deshpandeunmake}
\bibfield{author}{\bibinfo{person}{Himani Deshpande}, \bibinfo{person}{Haruki Takahashi}, {and} \bibinfo{person}{Jeeeun Kim}.} \bibinfo{year}{2024}\natexlab{}.
\newblock \showarticletitle{Unmake to Remake: Materiality-Driven Rapid Prototyping}.
\newblock \bibinfo{journal}{\emph{ACM Trans. Comput.-Hum. Interact.}} \bibinfo{volume}{31}, \bibinfo{number}{6}, Article \bibinfo{articleno}{78} (\bibinfo{date}{Dec.} \bibinfo{year}{2024}), \bibinfo{numpages}{31}~pages.
\newblock
\showISSN{1073-0516}
\urldef\tempurl%
\url{https://doi.org/10.1145/3685270}
\showDOI{\tempurl}


\bibitem[\protect\citeauthoryear{Dickey}{Dickey}{2017}]%
        {dickey2017stretchable}
\bibfield{author}{\bibinfo{person}{Michael~D. Dickey}.} \bibinfo{year}{2017}\natexlab{}.
\newblock \showarticletitle{Stretchable and Soft Electronics Using Liquid Metals}.
\newblock \bibinfo{journal}{\emph{Advanced Materials}} \bibinfo{volume}{29}, \bibinfo{number}{27} (\bibinfo{year}{2017}), \bibinfo{pages}{1606425}.
\newblock


\bibitem[\protect\citeauthoryear{Dickey, Chiechi, Larsen, Weiss, Weitz, and Whitesides}{Dickey et~al\mbox{.}}{2008}]%
        {dickey2008eutectic}
\bibfield{author}{\bibinfo{person}{Michael~D Dickey}, \bibinfo{person}{Ryan~C Chiechi}, \bibinfo{person}{Ryan~J Larsen}, \bibinfo{person}{Emily~A Weiss}, \bibinfo{person}{David~A Weitz}, {and} \bibinfo{person}{George~M Whitesides}.} \bibinfo{year}{2008}\natexlab{}.
\newblock \showarticletitle{Eutectic gallium-indium (EGaIn): a liquid metal alloy for the formation of stable structures in microchannels at room temperature}.
\newblock \bibinfo{journal}{\emph{Advanced functional materials}} \bibinfo{volume}{18}, \bibinfo{number}{7} (\bibinfo{year}{2008}), \bibinfo{pages}{1097--1104}.
\newblock


\bibitem[\protect\citeauthoryear{{ecoinvent Association}}{{ecoinvent Association}}{[n.d.]}]%
        {ecoinvent3.10}
\bibfield{author}{\bibinfo{person}{{ecoinvent Association}}.} \bibinfo{year}{[n.d.]}\natexlab{}.
\newblock \bibinfo{title}{{ecoinvent v3.10}: Life Cycle Inventory (LCI) Database}.
\newblock \bibinfo{howpublished}{\url{https://ecoinvent.org/ecoinvent-v3-10/}}.
\newblock
\newblock
\shownote{Accessed: April 06, 2025.}


\bibitem[\protect\citeauthoryear{electronicsworkshops}{electronicsworkshops}{2024}]%
        {MX1508MotorDriverModule}
\bibfield{author}{\bibinfo{person}{electronicsworkshops}.} \bibinfo{year}{2024}\natexlab{}.
\newblock \bibinfo{title}{MX1508 motor driver module}.
\newblock \bibinfo{howpublished}{\url{https://hackaday.io/project/197831-mx1508-motor-driver-module}}.
\newblock
\newblock
\shownote{Accessed: July 09, 2025.}


\bibitem[\protect\citeauthoryear{{Elmer's}}{{Elmer's}}{[n.d.]}]%
        {ElmersPVA}
\bibfield{author}{\bibinfo{person}{{Elmer's}}.} \bibinfo{year}{[n.d.]}\natexlab{}.
\newblock \bibinfo{title}{Liquid School Glue}.
\newblock \bibinfo{howpublished}{\url{https://www.elmers.com/glue/liquid-school-glue/}}.
\newblock
\newblock
\shownote{Accessed: April 06, 2025.}


\bibitem[\protect\citeauthoryear{Fassler and Majidi}{Fassler and Majidi}{2015}]%
        {fassler2015liquid}
\bibfield{author}{\bibinfo{person}{Andrew Fassler} {and} \bibinfo{person}{Carmel Majidi}.} \bibinfo{year}{2015}\natexlab{}.
\newblock \showarticletitle{Liquid-phase metal inclusions for a conductive polymer composite}.
\newblock \bibinfo{journal}{\emph{Adv. Mater}} \bibinfo{volume}{27}, \bibinfo{number}{11} (\bibinfo{year}{2015}), \bibinfo{pages}{1928--1932}.
\newblock


\bibitem[\protect\citeauthoryear{{Filabot}}{{Filabot}}{2024}]%
        {filabotEX2}
\bibfield{author}{\bibinfo{person}{{Filabot}}.} \bibinfo{year}{2024}\natexlab{}.
\newblock \bibinfo{title}{Filabot Original EX2}.
\newblock \bibinfo{howpublished}{\url{https://www.filabot.com/collections/filabot-core/products/filabot-original-ex2}}.
\newblock
\newblock
\shownote{Accessed: April 06, 2025.}


\bibitem[\protect\citeauthoryear{{FreeCAD Community}}{{FreeCAD Community}}{[n.d.]}]%
        {freecad}
\bibfield{author}{\bibinfo{person}{{FreeCAD Community}}.} \bibinfo{year}{[n.d.]}\natexlab{}.
\newblock \bibinfo{title}{FreeCAD: Open-Source Parametric 3D CAD Modeler}.
\newblock \bibinfo{howpublished}{\url{https://www.freecad.org/}}.
\newblock
\newblock
\shownote{Accessed: April 06, 2025.}


\bibitem[\protect\citeauthoryear{{GreenDelta GmbH}}{{GreenDelta GmbH}}{[n.d.]}]%
        {openLCA}
\bibfield{author}{\bibinfo{person}{{GreenDelta GmbH}}.} \bibinfo{year}{[n.d.]}\natexlab{}.
\newblock \bibinfo{title}{openLCA: Free and Open Source Life Cycle Assessment Software}.
\newblock \bibinfo{howpublished}{\url{https://www.openlca.org/}}.
\newblock
\newblock
\shownote{Accessed: April 06, 2025.}


\bibitem[\protect\citeauthoryear{Gu, Breen, Hu, Zhu, Tao, Van~de Zande, Wang, Zhang, and Yao}{Gu et~al\mbox{.}}{2019}]%
        {Geodesy}
\bibfield{author}{\bibinfo{person}{Jianzhe Gu}, \bibinfo{person}{David~E. Breen}, \bibinfo{person}{Jenny Hu}, \bibinfo{person}{Lifeng Zhu}, \bibinfo{person}{Ye Tao}, \bibinfo{person}{Tyson Van~de Zande}, \bibinfo{person}{Guanyun Wang}, \bibinfo{person}{Yongjie~Jessica Zhang}, {and} \bibinfo{person}{Lining Yao}.} \bibinfo{year}{2019}\natexlab{}.
\newblock \showarticletitle{Geodesy: Self-rising 2.5D Tiles by Printing along 2D Geodesic Closed Path}. In \bibinfo{booktitle}{\emph{Proceedings of the 2019 CHI Conference on Human Factors in Computing Systems}} (Glasgow, Scotland Uk) \emph{(\bibinfo{series}{CHI '19})}. \bibinfo{publisher}{Association for Computing Machinery}, \bibinfo{address}{New York, NY, USA}, \bibinfo{pages}{1–10}.
\newblock
\showISBNx{9781450359702}
\urldef\tempurl%
\url{https://doi.org/10.1145/3290605.3300267}
\showDOI{\tempurl}


\bibitem[\protect\citeauthoryear{Guna, Murugesan, Basavarajaiah, Ilangovan, Olivera, Krishna, and Reddy}{Guna et~al\mbox{.}}{2016}]%
        {guna2016plant}
\bibfield{author}{\bibinfo{person}{Vijay~Kumar Guna}, \bibinfo{person}{Geethapriya Murugesan}, \bibinfo{person}{Bhuvaneswari~Hulikal Basavarajaiah}, \bibinfo{person}{Manikandan Ilangovan}, \bibinfo{person}{Sharon Olivera}, \bibinfo{person}{Venkatesh Krishna}, {and} \bibinfo{person}{Narendra Reddy}.} \bibinfo{year}{2016}\natexlab{}.
\newblock \showarticletitle{Plant-based completely biodegradable printed circuit boards}.
\newblock \bibinfo{journal}{\emph{IEEE Transactions on Electron Devices}} \bibinfo{volume}{63}, \bibinfo{number}{12} (\bibinfo{year}{2016}), \bibinfo{pages}{4893--4898}.
\newblock


\bibitem[\protect\citeauthoryear{Guridi, Pouta, Hokkanen, and Jaiswal}{Guridi et~al\mbox{.}}{2023}]%
        {Cellulose-Based_Optical_Textile_Sensors}
\bibfield{author}{\bibinfo{person}{Sof\'{\i}a Guridi}, \bibinfo{person}{Emmi Pouta}, \bibinfo{person}{Ari Hokkanen}, {and} \bibinfo{person}{Aayush Jaiswal}.} \bibinfo{year}{2023}\natexlab{}.
\newblock \showarticletitle{LIGHT TISSUE: Development of Cellulose-Based Optical Textile Sensors}. In \bibinfo{booktitle}{\emph{Proceedings of the Seventeenth International Conference on Tangible, Embedded, and Embodied Interaction}} (Warsaw, Poland) \emph{(\bibinfo{series}{TEI '23})}. \bibinfo{publisher}{Association for Computing Machinery}, \bibinfo{address}{New York, NY, USA}, Article \bibinfo{articleno}{27}, \bibinfo{numpages}{14}~pages.
\newblock
\showISBNx{9781450399777}
\urldef\tempurl%
\url{https://doi.org/10.1145/3569009.3572798}
\showDOI{\tempurl}


\bibitem[\protect\citeauthoryear{Hauschild, Rosenbaum, Olsen, et~al\mbox{.}}{Hauschild et~al\mbox{.}}{2018}]%
        {hauschild2018life}
\bibfield{author}{\bibinfo{person}{Michael~Z Hauschild}, \bibinfo{person}{Ralph~K Rosenbaum}, \bibinfo{person}{Stig~Irving Olsen}, {et~al\mbox{.}}} \bibinfo{year}{2018}\natexlab{}.
\newblock \bibinfo{booktitle}{\emph{Life Cycle Assessment: Theory and Practice}}.
\newblock \bibinfo{publisher}{Springer International Publishing}, \bibinfo{address}{Cham, Switzerland}.
\newblock
\showISBNx{978-3-319-56475-3}
\urldef\tempurl%
\url{https://doi.org/10.1007/978-3-319-56475-3}
\showDOI{\tempurl}


\bibitem[\protect\citeauthoryear{He, Wittkopf, Jun, Erickson, and Ballagas}{He et~al\mbox{.}}{2022}]%
        {ModElec}
\bibfield{author}{\bibinfo{person}{Liang He}, \bibinfo{person}{Jarrid~A. Wittkopf}, \bibinfo{person}{Ji~Won Jun}, \bibinfo{person}{Kris Erickson}, {and} \bibinfo{person}{Rafael~Tico Ballagas}.} \bibinfo{year}{2022}\natexlab{}.
\newblock \showarticletitle{ModElec: A Design Tool for Prototyping Physical Computing Devices Using Conductive 3D Printing}.
\newblock \bibinfo{journal}{\emph{Proc. ACM Interact. Mob. Wearable Ubiquitous Technol.}} \bibinfo{volume}{5}, \bibinfo{number}{4}, Article \bibinfo{articleno}{159} (\bibinfo{date}{Dec.} \bibinfo{year}{2022}), \bibinfo{numpages}{20}~pages.
\newblock
\urldef\tempurl%
\url{https://doi.org/10.1145/3495000}
\showDOI{\tempurl}


\bibitem[\protect\citeauthoryear{Hillaire, Dickey, and Daniels}{Hillaire et~al\mbox{.}}{2020}]%
        {Hillaire2020MarangoniFI}
\bibfield{author}{\bibinfo{person}{Keith~D. Hillaire}, \bibinfo{person}{Michael~David Dickey}, {and} \bibinfo{person}{Karen~E. Daniels}.} \bibinfo{year}{2020}\natexlab{}.
\newblock \showarticletitle{Marangoni Fingering Instabilities in Oxidizing Liquid Metals}.
\newblock \bibinfo{journal}{\emph{arXiv: Fluid Dynamics}} (\bibinfo{year}{2020}).
\newblock
\urldef\tempurl%
\url{https://api.semanticscholar.org/CorpusID:227054206}
\showURL{%
\tempurl}


\bibitem[\protect\citeauthoryear{Huang, Liu, Hwang, Kang, Patnaik, Cortes, and Rogers}{Huang et~al\mbox{.}}{2014}]%
        {Transient_PCB}
\bibfield{author}{\bibinfo{person}{Xian Huang}, \bibinfo{person}{Yuhao Liu}, \bibinfo{person}{Suk-Won Hwang}, \bibinfo{person}{Seung-Kyun Kang}, \bibinfo{person}{Dwipayan Patnaik}, \bibinfo{person}{Jonathan~Fajardo Cortes}, {and} \bibinfo{person}{John~A. Rogers}.} \bibinfo{year}{2014}\natexlab{}.
\newblock \showarticletitle{Biodegradable Materials for Multilayer Transient Printed Circuit Boards}.
\newblock \bibinfo{journal}{\emph{Advanced Materials}} \bibinfo{volume}{26}, \bibinfo{number}{43} (\bibinfo{year}{2014}), \bibinfo{pages}{7371--7377}.
\newblock
\urldef\tempurl%
\url{https://onlinelibrary.wiley.com/doi/abs/10.1002/adma.201403164}
\showURL{%
\tempurl}


\bibitem[\protect\citeauthoryear{Hwang, Tao, Kim, Cheng, Song, Rill, Brenckle, Panilaitis, Won, Kim, Song, Yu, Ameen, Li, Su, Yang, Kaplan, Zakin, Slepian, Huang, Omenetto, and Rogers}{Hwang et~al\mbox{.}}{2012}]%
        {Transient_Electronics}
\bibfield{author}{\bibinfo{person}{Suk-Won Hwang}, \bibinfo{person}{Hu Tao}, \bibinfo{person}{Dae-Hyeong Kim}, \bibinfo{person}{Huanyu Cheng}, \bibinfo{person}{Jun-Kyul Song}, \bibinfo{person}{Elliott Rill}, \bibinfo{person}{Mark~A. Brenckle}, \bibinfo{person}{Bruce Panilaitis}, \bibinfo{person}{Sang~Min Won}, \bibinfo{person}{Yun-Soung Kim}, \bibinfo{person}{Young~Min Song}, \bibinfo{person}{Ki~Jun Yu}, \bibinfo{person}{Abid Ameen}, \bibinfo{person}{Rui Li}, \bibinfo{person}{Yewang Su}, \bibinfo{person}{Miaomiao Yang}, \bibinfo{person}{David~L. Kaplan}, \bibinfo{person}{Mitchell~R. Zakin}, \bibinfo{person}{Marvin~J. Slepian}, \bibinfo{person}{Yonggang Huang}, \bibinfo{person}{Fiorenzo~G. Omenetto}, {and} \bibinfo{person}{John~A. Rogers}.} \bibinfo{year}{2012}\natexlab{}.
\newblock \showarticletitle{A Physically Transient Form of Silicon Electronics}.
\newblock \bibinfo{journal}{\emph{Science}} \bibinfo{volume}{337}, \bibinfo{number}{6102} (\bibinfo{year}{2012}), \bibinfo{pages}{1640--1644}.
\newblock
\urldef\tempurl%
\url{https://doi.org/10.1126/science.1226325}
\showDOI{\tempurl}


\bibitem[\protect\citeauthoryear{Insights}{Insights}{2025}]%
        {deloitte2025}
\bibfield{author}{\bibinfo{person}{Deloitte Insights}.} \bibinfo{year}{2025}\natexlab{}.
\newblock \bibinfo{title}{2025 Global Semiconductor Industry Outlook}.
\newblock \bibinfo{howpublished}{\url{https://www2.deloitte.com/us/en/insights/industry/technology/technology-media-telecom-outlooks/semiconductor-industry-outlook.html}}.
\newblock
\newblock
\shownote{Accessed: April 06, 2025.}


\bibitem[\protect\citeauthoryear{{Institute of Environmental Sciences (CML), Leiden University}}{{Institute of Environmental Sciences (CML), Leiden University}}{2016}]%
        {CMLv4.8}
\bibfield{author}{\bibinfo{person}{{Institute of Environmental Sciences (CML), Leiden University}}.} \bibinfo{year}{2016}\natexlab{}.
\newblock \bibinfo{title}{CML‑IA Characterisation Factors for Life Cycle Impact Assessment}.
\newblock \bibinfo{howpublished}{\url{https://www.universiteitleiden.nl/en/research/research-output/science/cml-ia-characterisation-factors}}.
\newblock
\newblock
\shownote{Accessed: April 06, 2025.}


\bibitem[\protect\citeauthoryear{{International Electrotechnical Commission}}{{International Electrotechnical Commission}}{2013}]%
        {iec60529}
\bibfield{author}{\bibinfo{person}{{International Electrotechnical Commission}}.} \bibinfo{year}{2013}\natexlab{}.
\newblock \bibinfo{title}{{IEC 60529: Degrees of Protection Provided by Enclosures (IP Code)}}.
\newblock
\newblock
\urldef\tempurl%
\url{https://www.iec.ch/ip-ratings}
\showURL{%
\tempurl}


\bibitem[\protect\citeauthoryear{Ishii, Kato, Ikematsu, Kawahara, and Siio}{Ishii et~al\mbox{.}}{2021}]%
        {Wooden_Circuit}
\bibfield{author}{\bibinfo{person}{Ayaka Ishii}, \bibinfo{person}{Kunihiro Kato}, \bibinfo{person}{Kaori Ikematsu}, \bibinfo{person}{Yoshihiro Kawahara}, {and} \bibinfo{person}{Itiro Siio}.} \bibinfo{year}{2021}\natexlab{}.
\newblock \showarticletitle{Fabricating Wooden Circuit Boards by Laser Beam Machining}. In \bibinfo{booktitle}{\emph{Adjunct Proceedings of the 34th Annual ACM Symposium on User Interface Software and Technology}} (Virtual Event, USA) \emph{(\bibinfo{series}{UIST '21 Adjunct})}. \bibinfo{publisher}{Association for Computing Machinery}, \bibinfo{address}{New York, NY, USA}, \bibinfo{pages}{109–111}.
\newblock
\showISBNx{9781450386555}
\urldef\tempurl%
\url{https://doi.org/10.1145/3474349.3480191}
\showDOI{\tempurl}


\bibitem[\protect\citeauthoryear{{Jiva Materials}}{{Jiva Materials}}{[n.d.]}]%
        {jiva}
\bibfield{author}{\bibinfo{person}{{Jiva Materials}}.} \bibinfo{year}{[n.d.]}\natexlab{}.
\newblock \bibinfo{title}{Jiva Materials}.
\newblock \bibinfo{howpublished}{\url{https://www.jivamaterials.com/}}.
\newblock
\newblock
\shownote{Accessed: April 06, 2025.}


\bibitem[\protect\citeauthoryear{Kawahara, Hodges, Cook, Zhang, and Abowd}{Kawahara et~al\mbox{.}}{2013}]%
        {Instant_Inkjet_Circuits}
\bibfield{author}{\bibinfo{person}{Yoshihiro Kawahara}, \bibinfo{person}{Steve Hodges}, \bibinfo{person}{Benjamin~S. Cook}, \bibinfo{person}{Cheng Zhang}, {and} \bibinfo{person}{Gregory~D. Abowd}.} \bibinfo{year}{2013}\natexlab{}.
\newblock \showarticletitle{Instant Inkjet Circuits: Lab-Based Inkjet Printing to Support Rapid Prototyping of UbiComp Devices}. In \bibinfo{booktitle}{\emph{Proceedings of the 2013 ACM International Joint Conference on Pervasive and Ubiquitous Computing}} (Zurich, Switzerland) \emph{(\bibinfo{series}{UbiComp '13})}. \bibinfo{publisher}{Association for Computing Machinery}, \bibinfo{address}{New York, NY, USA}, \bibinfo{pages}{363–372}.
\newblock
\showISBNx{9781450317702}
\urldef\tempurl%
\url{https://doi.org/10.1145/2493432.2493486}
\showDOI{\tempurl}


\bibitem[\protect\citeauthoryear{{KiCad Project}}{{KiCad Project}}{[n.d.]}]%
        {kicad}
\bibfield{author}{\bibinfo{person}{{KiCad Project}}.} \bibinfo{year}{[n.d.]}\natexlab{}.
\newblock \bibinfo{title}{KiCad EDA: A Cross-Platform Open-Source Electronics Design Automation Suite}.
\newblock \bibinfo{howpublished}{\url{https://kicad.org/}}.
\newblock
\newblock
\shownote{Accessed: April 06, 2025.}


\bibitem[\protect\citeauthoryear{Kiddee, Naidu, and Wong}{Kiddee et~al\mbox{.}}{2013}]%
        {Kiddee2013}
\bibfield{author}{\bibinfo{person}{Peeranart Kiddee}, \bibinfo{person}{Ravi Naidu}, {and} \bibinfo{person}{Ming~H. Wong}.} \bibinfo{year}{2013}\natexlab{}.
\newblock \showarticletitle{Electronic waste management approaches: An overview}.
\newblock \bibinfo{journal}{\emph{Waste Management}} \bibinfo{volume}{33}, \bibinfo{number}{5} (\bibinfo{year}{2013}), \bibinfo{pages}{1237--1250}.
\newblock


\bibitem[\protect\citeauthoryear{Kim and Paulos}{Kim and Paulos}{2011}]%
        {Creative_Reuse}
\bibfield{author}{\bibinfo{person}{Sunyoung Kim} {and} \bibinfo{person}{Eric Paulos}.} \bibinfo{year}{2011}\natexlab{}.
\newblock \showarticletitle{Practices in the Creative Reuse of E-Waste}. In \bibinfo{booktitle}{\emph{Proceedings of the SIGCHI Conference on Human Factors in Computing Systems}} (Vancouver, BC, Canada) \emph{(\bibinfo{series}{CHI '11})}. \bibinfo{publisher}{Association for Computing Machinery}, \bibinfo{address}{New York, NY, USA}, \bibinfo{pages}{2395–2404}.
\newblock
\showISBNx{9781450302289}
\urldef\tempurl%
\url{https://doi.org/10.1145/1978942.1979292}
\showDOI{\tempurl}


\bibitem[\protect\citeauthoryear{Koelle, Nicolae, Nittala, Teyssier, and Steimle}{Koelle et~al\mbox{.}}{2022}]%
        {Bioplastics}
\bibfield{author}{\bibinfo{person}{Marion Koelle}, \bibinfo{person}{Madalina Nicolae}, \bibinfo{person}{Aditya~Shekhar Nittala}, \bibinfo{person}{Marc Teyssier}, {and} \bibinfo{person}{J\"{u}rgen Steimle}.} \bibinfo{year}{2022}\natexlab{}.
\newblock \showarticletitle{Prototyping Soft Devices with Interactive Bioplastics}. In \bibinfo{booktitle}{\emph{Proceedings of the 35th Annual ACM Symposium on User Interface Software and Technology}} (Bend, OR, USA) \emph{(\bibinfo{series}{UIST '22})}. \bibinfo{publisher}{Association for Computing Machinery}, \bibinfo{address}{New York, NY, USA}, Article \bibinfo{articleno}{19}, \bibinfo{numpages}{16}~pages.
\newblock
\showISBNx{9781450393201}
\urldef\tempurl%
\url{https://doi.org/10.1145/3526113.3545623}
\showDOI{\tempurl}


\bibitem[\protect\citeauthoryear{Lazaro~Vasquez, Alistar, Devendorf, and Rivera}{Lazaro~Vasquez et~al\mbox{.}}{2024}]%
        {spinning}
\bibfield{author}{\bibinfo{person}{Eldy~S. Lazaro~Vasquez}, \bibinfo{person}{Mirela Alistar}, \bibinfo{person}{Laura Devendorf}, {and} \bibinfo{person}{Michael~L. Rivera}.} \bibinfo{year}{2024}\natexlab{}.
\newblock \showarticletitle{Desktop Biofibers Spinning: An Open-Source Machine for Exploring Biobased Fibers and Their Application Towards Sustainable Smart Textile Design}. In \bibinfo{booktitle}{\emph{Proceedings of the CHI Conference on Human Factors in Computing Systems}} (Honolulu, HI, USA) \emph{(\bibinfo{series}{CHI '24})}. \bibinfo{publisher}{Association for Computing Machinery}, \bibinfo{address}{New York, NY, USA}, Article \bibinfo{articleno}{856}, \bibinfo{numpages}{18}~pages.
\newblock
\showISBNx{9798400703300}
\urldef\tempurl%
\url{https://doi.org/10.1145/3613904.3642387}
\showDOI{\tempurl}


\bibitem[\protect\citeauthoryear{Lu, Desta, Wu, Nith, Passananti, and Lopes}{Lu et~al\mbox{.}}{2023}]%
        {ecoeda}
\bibfield{author}{\bibinfo{person}{Jasmine Lu}, \bibinfo{person}{Beza Desta}, \bibinfo{person}{K.~D. Wu}, \bibinfo{person}{Romain Nith}, \bibinfo{person}{Joyce~E Passananti}, {and} \bibinfo{person}{Pedro Lopes}.} \bibinfo{year}{2023}\natexlab{}.
\newblock \showarticletitle{EcoEDA: Recycling E-Waste During Electronics Design}. In \bibinfo{booktitle}{\emph{Proceedings of the 36th Annual ACM Symposium on User Interface Software and Technology}} (San Francisco, CA, USA) \emph{(\bibinfo{series}{UIST '23})}. \bibinfo{publisher}{Association for Computing Machinery}, \bibinfo{address}{New York, NY, USA}, Article \bibinfo{articleno}{30}, \bibinfo{numpages}{14}~pages.
\newblock
\showISBNx{9798400701320}
\urldef\tempurl%
\url{https://doi.org/10.1145/3586183.3606745}
\showDOI{\tempurl}


\bibitem[\protect\citeauthoryear{Lu, Rishitha~Boddu, and Lopes}{Lu et~al\mbox{.}}{2025}]%
        {ProtoPCB}
\bibfield{author}{\bibinfo{person}{Jasmine Lu}, \bibinfo{person}{Sai Rishitha~Boddu}, {and} \bibinfo{person}{Pedro Lopes}.} \bibinfo{year}{2025}\natexlab{}.
\newblock \showarticletitle{ProtoPCB: Reclaiming Printed Circuit Board E-waste as Prototyping Material}. In \bibinfo{booktitle}{\emph{Proceedings of the 2025 CHI Conference on Human Factors in Computing Systems}} \emph{(\bibinfo{series}{CHI '25})}. \bibinfo{publisher}{Association for Computing Machinery}, \bibinfo{address}{New York, NY, USA}.
\newblock
\urldef\tempurl%
\url{https://doi.org/10.1145/3613904.3642765}
\showDOI{\tempurl}


\bibitem[\protect\citeauthoryear{Maurice, Dinh, Charpentier, Brambilla, and Gabriel}{Maurice et~al\mbox{.}}{2021}]%
        {su131810357}
\bibfield{author}{\bibinfo{person}{Ange~A. Maurice}, \bibinfo{person}{Khang~Ngoc Dinh}, \bibinfo{person}{Nicolas~M. Charpentier}, \bibinfo{person}{Andrea Brambilla}, {and} \bibinfo{person}{Jean-Christophe~P. Gabriel}.} \bibinfo{year}{2021}\natexlab{}.
\newblock \showarticletitle{Dismantling of Printed Circuit Boards Enabling Electronic Components Sorting and Their Subsequent Treatment Open Improved Elemental Sustainability Opportunities}.
\newblock \bibinfo{journal}{\emph{Sustainability}} \bibinfo{volume}{13}, \bibinfo{number}{18} (\bibinfo{year}{2021}).
\newblock
\showISSN{2071-1050}
\urldef\tempurl%
\url{https://doi.org/10.3390/su131810357}
\showDOI{\tempurl}


\bibitem[\protect\citeauthoryear{Momeni, {M.Mehdi Hassani.N}, Liu, and Ni}{Momeni et~al\mbox{.}}{2017}]%
        {MOMENI201742}
\bibfield{author}{\bibinfo{person}{Farhang Momeni}, \bibinfo{person}{Seyed {M.Mehdi Hassani.N}}, \bibinfo{person}{Xun Liu}, {and} \bibinfo{person}{Jun Ni}.} \bibinfo{year}{2017}\natexlab{}.
\newblock \showarticletitle{A review of 4D printing}.
\newblock \bibinfo{journal}{\emph{Materials \& Design}}  \bibinfo{volume}{122} (\bibinfo{year}{2017}), \bibinfo{pages}{42--79}.
\newblock
\showISSN{0264-1275}
\urldef\tempurl%
\url{https://doi.org/10.1016/j.matdes.2017.02.068}
\showDOI{\tempurl}


\bibitem[\protect\citeauthoryear{Murer, Vallg\r{a}rda, Jacobsson, and Tscheligi}{Murer et~al\mbox{.}}{2015}]%
        {murer2015crafting}
\bibfield{author}{\bibinfo{person}{Martin Murer}, \bibinfo{person}{Anna Vallg\r{a}rda}, \bibinfo{person}{Mattias Jacobsson}, {and} \bibinfo{person}{Manfred Tscheligi}.} \bibinfo{year}{2015}\natexlab{}.
\newblock \showarticletitle{Un-Crafting: Exploring Tangible Practices for Deconstruction in Interactive System Design}. In \bibinfo{booktitle}{\emph{Proceedings of the Ninth International Conference on Tangible, Embedded, and Embodied Interaction}} (Stanford, California, USA) \emph{(\bibinfo{series}{TEI '15})}. \bibinfo{publisher}{Association for Computing Machinery}, \bibinfo{address}{New York, NY, USA}, \bibinfo{pages}{469–472}.
\newblock
\showISBNx{9781450333054}
\urldef\tempurl%
\url{https://doi.org/10.1145/2677199.2683582}
\showDOI{\tempurl}


\bibitem[\protect\citeauthoryear{Nagels, Ramakers, Luyten, and Deferme}{Nagels et~al\mbox{.}}{2018}]%
        {Silicone_Devices}
\bibfield{author}{\bibinfo{person}{Steven Nagels}, \bibinfo{person}{Raf Ramakers}, \bibinfo{person}{Kris Luyten}, {and} \bibinfo{person}{Wim Deferme}.} \bibinfo{year}{2018}\natexlab{}.
\newblock \showarticletitle{Silicone Devices: A Scalable DIY Approach for Fabricating Self-Contained Multi-Layered Soft Circuits using Microfluidics}. In \bibinfo{booktitle}{\emph{Proceedings of the 2018 CHI Conference on Human Factors in Computing Systems}} (Montreal QC, Canada) \emph{(\bibinfo{series}{CHI '18})}. \bibinfo{publisher}{Association for Computing Machinery}, \bibinfo{address}{New York, NY, USA}, \bibinfo{pages}{1–13}.
\newblock
\showISBNx{9781450356206}
\urldef\tempurl%
\url{https://doi.org/10.1145/3173574.3173762}
\showDOI{\tempurl}


\bibitem[\protect\citeauthoryear{Park, Ryu, P{\'o}czos, Shin, Lee, Chu, Rogers, Shin, Kim, et~al\mbox{.}}{Park et~al\mbox{.}}{2010}]%
        {park2010hyperelastic}
\bibfield{author}{\bibinfo{person}{Y-L Park}, \bibinfo{person}{Sechang Ryu}, \bibinfo{person}{Barnab{\'a}s P{\'o}czos}, \bibinfo{person}{Kyeong-Sik Shin}, \bibinfo{person}{Bongsoo Lee}, \bibinfo{person}{Ung~B. Chu}, \bibinfo{person}{John~A. Rogers}, \bibinfo{person}{Sooyoung Shin}, \bibinfo{person}{Seon~Jeong Kim}, {et~al\mbox{.}}} \bibinfo{year}{2010}\natexlab{}.
\newblock \showarticletitle{Hyperelastic Pressure Sensing with a Liquid-Embedded Elastomer}.
\newblock \bibinfo{journal}{\emph{Journal of Micromechanics and Microengineering}} \bibinfo{volume}{20}, \bibinfo{number}{12} (\bibinfo{year}{2010}), \bibinfo{pages}{125029}.
\newblock


\bibitem[\protect\citeauthoryear{Pourjafarian, Yang, Lipton, Davaji, and Abowd}{Pourjafarian et~al\mbox{.}}{2025}]%
        {proform}
\bibfield{author}{\bibinfo{person}{Narjes Pourjafarian}, \bibinfo{person}{Zhenming Yang}, \bibinfo{person}{Jeffrey Lipton}, \bibinfo{person}{Benyamin Davaji}, {and} \bibinfo{person}{Gregory~D Abowd}.} \bibinfo{year}{2025}\natexlab{}.
\newblock \showarticletitle{ProForm: Solder-Free Circuit Assembly Using Thermoforming}. In \bibinfo{booktitle}{\emph{Proceedings of the 38th Annual ACM Symposium on User Interface Software and Technologys}} (Busan, Republic of Korea) \emph{(\bibinfo{series}{UIST '25})}. \bibinfo{publisher}{Association for Computing Machinery}, \bibinfo{address}{New York, NY, USA}.
\newblock
\urldef\tempurl%
\url{https://doi.org/10.1145/3746059.3747628}
\showDOI{\tempurl}


\bibitem[\protect\citeauthoryear{Rivera, Bae, and Hudson}{Rivera et~al\mbox{.}}{2023}]%
        {Coffee_Grounds}
\bibfield{author}{\bibinfo{person}{Michael~L. Rivera}, \bibinfo{person}{S.~Sandra Bae}, {and} \bibinfo{person}{Scott~E. Hudson}.} \bibinfo{year}{2023}\natexlab{}.
\newblock \showarticletitle{Designing a Sustainable Material for 3D Printing with Spent Coffee Grounds}. In \bibinfo{booktitle}{\emph{Proceedings of the 2023 ACM Designing Interactive Systems Conference}} (Pittsburgh, PA, USA) \emph{(\bibinfo{series}{DIS '23})}. \bibinfo{publisher}{Association for Computing Machinery}, \bibinfo{address}{New York, NY, USA}, \bibinfo{pages}{294–311}.
\newblock
\showISBNx{9781450398930}
\urldef\tempurl%
\url{https://doi.org/10.1145/3563657.3595983}
\showDOI{\tempurl}


\bibitem[\protect\citeauthoryear{Savage, Schmidt, Grossman, Fitzmaurice, and Hartmann}{Savage et~al\mbox{.}}{2014}]%
        {A_series_of_tubes}
\bibfield{author}{\bibinfo{person}{Valkyrie Savage}, \bibinfo{person}{Ryan Schmidt}, \bibinfo{person}{Tovi Grossman}, \bibinfo{person}{George Fitzmaurice}, {and} \bibinfo{person}{Bj\"{o}rn Hartmann}.} \bibinfo{year}{2014}\natexlab{}.
\newblock \showarticletitle{A series of tubes: adding interactivity to 3D prints using internal pipes}. In \bibinfo{booktitle}{\emph{Proceedings of the 27th Annual ACM Symposium on User Interface Software and Technology}} (Honolulu, Hawaii, USA) \emph{(\bibinfo{series}{UIST '14})}. \bibinfo{publisher}{Association for Computing Machinery}, \bibinfo{address}{New York, NY, USA}, \bibinfo{pages}{3–12}.
\newblock
\showISBNx{9781450330695}
\urldef\tempurl%
\url{https://doi.org/10.1145/2642918.2647374}
\showDOI{\tempurl}


\bibitem[\protect\citeauthoryear{Song, Bell, Deshpande, Mandel, Wun, Alistar, Buechley, Ju, Kim, Paulos, et~al\mbox{.}}{Song et~al\mbox{.}}{2024}]%
        {song2024sustainable}
\bibfield{author}{\bibinfo{person}{Katherine~W Song}, \bibinfo{person}{Fiona Bell}, \bibinfo{person}{Himani Deshpande}, \bibinfo{person}{Ilan Mandel}, \bibinfo{person}{Tiffany Wun}, \bibinfo{person}{Mirela Alistar}, \bibinfo{person}{Leah Buechley}, \bibinfo{person}{Wendy Ju}, \bibinfo{person}{Jeeeun Kim}, \bibinfo{person}{Eric Paulos}, {et~al\mbox{.}}} \bibinfo{year}{2024}\natexlab{}.
\newblock \showarticletitle{Sustainable Unmaking: Designing for Biodegradation, Decay, and Disassembly}. In \bibinfo{booktitle}{\emph{Extended Abstracts of the CHI Conference on Human Factors in Computing Systems}}. \bibinfo{pages}{1--7}.
\newblock


\bibitem[\protect\citeauthoryear{Song, Maheshwari, Gallo, Danielescu, and Paulos}{Song et~al\mbox{.}}{2022}]%
        {song2022towards}
\bibfield{author}{\bibinfo{person}{Katherine~W Song}, \bibinfo{person}{Aditi Maheshwari}, \bibinfo{person}{Eric~M Gallo}, \bibinfo{person}{Andreea Danielescu}, {and} \bibinfo{person}{Eric Paulos}.} \bibinfo{year}{2022}\natexlab{}.
\newblock \showarticletitle{Towards Decomposable Interactive Systems: Design of a Backyard-Degradable Wireless Heating Interface}. In \bibinfo{booktitle}{\emph{Proceedings of the 2022 CHI Conference on Human Factors in Computing Systems}} (New Orleans, LA, USA) \emph{(\bibinfo{series}{CHI '22})}. \bibinfo{publisher}{Association for Computing Machinery}, \bibinfo{address}{New York, NY, USA}, Article \bibinfo{articleno}{100}, \bibinfo{numpages}{12}~pages.
\newblock
\showISBNx{9781450391573}
\urldef\tempurl%
\url{https://doi.org/10.1145/3491102.3502007}
\showDOI{\tempurl}


\bibitem[\protect\citeauthoryear{Song and Paulos}{Song and Paulos}{2021}]%
        {unmaking}
\bibfield{author}{\bibinfo{person}{Katherine~W Song} {and} \bibinfo{person}{Eric Paulos}.} \bibinfo{year}{2021}\natexlab{}.
\newblock \showarticletitle{Unmaking: Enabling and Celebrating the Creative Material of Failure, Destruction, Decay, and Deformation}. In \bibinfo{booktitle}{\emph{Proceedings of the 2021 CHI Conference on Human Factors in Computing Systems}} \emph{(\bibinfo{series}{CHI '21})}. \bibinfo{publisher}{Association for Computing Machinery}, \bibinfo{address}{New York, NY, USA}, Article \bibinfo{articleno}{429}, \bibinfo{numpages}{12}~pages.
\newblock
\showISBNx{9781450380966}
\urldef\tempurl%
\url{https://doi.org/10.1145/3411764.3445529}
\showDOI{\tempurl}


\bibitem[\protect\citeauthoryear{Song and Paulos}{Song and Paulos}{2023}]%
        {song2023vim}
\bibfield{author}{\bibinfo{person}{Katherine~W Song} {and} \bibinfo{person}{Eric Paulos}.} \bibinfo{year}{2023}\natexlab{}.
\newblock \showarticletitle{Vim: Customizable, Decomposable Electrical Energy Storage}. In \bibinfo{booktitle}{\emph{Proceedings of the 2023 CHI Conference on Human Factors in Computing Systems}} (Hamburg, Germany) \emph{(\bibinfo{series}{CHI '23})}. \bibinfo{publisher}{Association for Computing Machinery}, \bibinfo{address}{New York, NY, USA}, Article \bibinfo{articleno}{180}, \bibinfo{numpages}{18}~pages.
\newblock
\showISBNx{9781450394215}
\urldef\tempurl%
\url{https://doi.org/10.1145/3544548.3581110}
\showDOI{\tempurl}


\bibitem[\protect\citeauthoryear{Teng, Li, Huang, Li, Hu, Zhou, and Wang}{Teng et~al\mbox{.}}{2023}]%
        {teng2023fully}
\bibfield{author}{\bibinfo{person}{Long Teng}, \bibinfo{person}{Li Li}, \bibinfo{person}{Jingxia Huang}, \bibinfo{person}{Shuai Li}, \bibinfo{person}{Renchao Hu}, \bibinfo{person}{Xuechang Zhou}, {and} \bibinfo{person}{Hong Wang}.} \bibinfo{year}{2023}\natexlab{}.
\newblock \showarticletitle{Fully Recyclable Liquid-Metal-Based Multi-Layer Thermally Triggered Transient Electronic Devices}.
\newblock \bibinfo{journal}{\emph{Advanced Materials Technologies}} \bibinfo{volume}{8}, \bibinfo{number}{4} (\bibinfo{year}{2023}), \bibinfo{pages}{2201031}.
\newblock


\bibitem[\protect\citeauthoryear{Teng, Ye, Handschuh-Wang, Zhou, Gan, and Zhou}{Teng et~al\mbox{.}}{2019}]%
        {teng2019liquid}
\bibfield{author}{\bibinfo{person}{Long Teng}, \bibinfo{person}{Shichao Ye}, \bibinfo{person}{Stephan Handschuh-Wang}, \bibinfo{person}{Xiaohu Zhou}, \bibinfo{person}{Tiansheng Gan}, {and} \bibinfo{person}{Xuechang Zhou}.} \bibinfo{year}{2019}\natexlab{}.
\newblock \showarticletitle{Liquid metal-based transient circuits for flexible and recyclable electronics}.
\newblock \bibinfo{journal}{\emph{Advanced Functional Materials}} \bibinfo{volume}{29}, \bibinfo{number}{11} (\bibinfo{year}{2019}), \bibinfo{pages}{1808739}.
\newblock


\bibitem[\protect\citeauthoryear{Tokuda, Sahoo, Jones, Subramanian, and Withana}{Tokuda et~al\mbox{.}}{2021}]%
        {Flowcuits}
\bibfield{author}{\bibinfo{person}{Yutaka Tokuda}, \bibinfo{person}{Deepak~Ranjan Sahoo}, \bibinfo{person}{Matt Jones}, \bibinfo{person}{Sriram Subramanian}, {and} \bibinfo{person}{Anusha Withana}.} \bibinfo{year}{2021}\natexlab{}.
\newblock \showarticletitle{Flowcuits: Crafting Tangible and Interactive Electrical Components with Liquid Metal Circuits}. In \bibinfo{booktitle}{\emph{Proceedings of the Fifteenth International Conference on Tangible, Embedded, and Embodied Interaction}} (Salzburg, Austria) \emph{(\bibinfo{series}{TEI '21})}. \bibinfo{publisher}{Association for Computing Machinery}, \bibinfo{address}{New York, NY, USA}, Article \bibinfo{articleno}{35}, \bibinfo{numpages}{11}~pages.
\newblock
\showISBNx{9781450382137}
\urldef\tempurl%
\url{https://doi.org/10.1145/3430524.3440654}
\showDOI{\tempurl}


\bibitem[\protect\citeauthoryear{Vasquez and Vega}{Vasquez and Vega}{2019a}]%
        {10.1145/3341162.3343808}
\bibfield{author}{\bibinfo{person}{Eldy S.~Lazaro Vasquez} {and} \bibinfo{person}{Katia Vega}.} \bibinfo{year}{2019}\natexlab{a}.
\newblock \showarticletitle{From plastic to biomaterials: prototyping DIY electronics with mycelium}. In \bibinfo{booktitle}{\emph{Adjunct Proceedings of the 2019 ACM International Joint Conference on Pervasive and Ubiquitous Computing and Proceedings of the 2019 ACM International Symposium on Wearable Computers}} (London, United Kingdom) \emph{(\bibinfo{series}{UbiComp/ISWC '19 Adjunct})}. \bibinfo{publisher}{Association for Computing Machinery}, \bibinfo{address}{New York, NY, USA}, \bibinfo{pages}{308–311}.
\newblock
\showISBNx{9781450368698}
\urldef\tempurl%
\url{https://doi.org/10.1145/3341162.3343808}
\showDOI{\tempurl}


\bibitem[\protect\citeauthoryear{Vasquez and Vega}{Vasquez and Vega}{2019b}]%
        {Myco-Accessories}
\bibfield{author}{\bibinfo{person}{Eldy S.~Lazaro Vasquez} {and} \bibinfo{person}{Katia Vega}.} \bibinfo{year}{2019}\natexlab{b}.
\newblock \showarticletitle{Myco-Accessories: Sustainable Wearables with Biodegradable Materials}. In \bibinfo{booktitle}{\emph{Proceedings of the 2019 ACM International Symposium on Wearable Computers}} (London, United Kingdom) \emph{(\bibinfo{series}{ISWC '19})}. \bibinfo{publisher}{Association for Computing Machinery}, \bibinfo{address}{New York, NY, USA}, \bibinfo{pages}{306–311}.
\newblock
\showISBNx{9781450368704}
\urldef\tempurl%
\url{https://doi.org/10.1145/3341163.3346938}
\showDOI{\tempurl}


\bibitem[\protect\citeauthoryear{Wall, Jacobson, Vogel, and Schneider}{Wall et~al\mbox{.}}{2021}]%
        {Scrappy}
\bibfield{author}{\bibinfo{person}{Ludwig~Wilhelm Wall}, \bibinfo{person}{Alec Jacobson}, \bibinfo{person}{Daniel Vogel}, {and} \bibinfo{person}{Oliver Schneider}.} \bibinfo{year}{2021}\natexlab{}.
\newblock \showarticletitle{Scrappy: Using Scrap Material as Infill to Make Fabrication More Sustainable}. In \bibinfo{booktitle}{\emph{Proceedings of the 2021 CHI Conference on Human Factors in Computing Systems}} (Yokohama, Japan) \emph{(\bibinfo{series}{CHI '21})}. \bibinfo{publisher}{Association for Computing Machinery}, \bibinfo{address}{New York, NY, USA}, Article \bibinfo{articleno}{665}, \bibinfo{numpages}{12}~pages.
\newblock
\showISBNx{9781450380966}
\urldef\tempurl%
\url{https://doi.org/10.1145/3411764.3445187}
\showDOI{\tempurl}


\bibitem[\protect\citeauthoryear{Wall, Schneider, and Vogel}{Wall et~al\mbox{.}}{2023}]%
        {Substiports}
\bibfield{author}{\bibinfo{person}{Ludwig~Wilhelm Wall}, \bibinfo{person}{Oliver Schneider}, {and} \bibinfo{person}{Daniel Vogel}.} \bibinfo{year}{2023}\natexlab{}.
\newblock \showarticletitle{Substiports: User-Inserted Ad Hoc Objects as Reusable Structural Support for Unmodified FDM 3D Printers}. In \bibinfo{booktitle}{\emph{Proceedings of the 36th Annual ACM Symposium on User Interface Software and Technology}} (San Francisco, CA, USA) \emph{(\bibinfo{series}{UIST '23})}. \bibinfo{publisher}{Association for Computing Machinery}, \bibinfo{address}{New York, NY, USA}, Article \bibinfo{articleno}{34}, \bibinfo{numpages}{20}~pages.
\newblock
\showISBNx{9798400701320}
\urldef\tempurl%
\url{https://doi.org/10.1145/3586183.3606718}
\showDOI{\tempurl}


\bibitem[\protect\citeauthoryear{Wang, Cheng, Do, Yang, Tao, Gu, An, and Yao}{Wang et~al\mbox{.}}{2018a}]%
        {Printed_Paper_Actuator}
\bibfield{author}{\bibinfo{person}{Guanyun Wang}, \bibinfo{person}{Tingyu Cheng}, \bibinfo{person}{Youngwook Do}, \bibinfo{person}{Humphrey Yang}, \bibinfo{person}{Ye Tao}, \bibinfo{person}{Jianzhe Gu}, \bibinfo{person}{Byoungkwon An}, {and} \bibinfo{person}{Lining Yao}.} \bibinfo{year}{2018}\natexlab{a}.
\newblock \showarticletitle{Printed Paper Actuator: A Low-Cost Reversible Actuation and Sensing Method for Shape Changing Interfaces}. In \bibinfo{booktitle}{\emph{Proceedings of the 2018 CHI Conference on Human Factors in Computing Systems}} (Montreal QC, Canada) \emph{(\bibinfo{series}{CHI '18})}. \bibinfo{publisher}{Association for Computing Machinery}, \bibinfo{address}{New York, NY, USA}, \bibinfo{pages}{1–12}.
\newblock
\showISBNx{9781450356206}
\urldef\tempurl%
\url{https://doi.org/10.1145/3173574.3174143}
\showDOI{\tempurl}


\bibitem[\protect\citeauthoryear{Wang, Yang, Yan, Gecer~Ulu, Tao, Gu, Kara, and Yao}{Wang et~al\mbox{.}}{2018b}]%
        {4DMesh}
\bibfield{author}{\bibinfo{person}{Guanyun Wang}, \bibinfo{person}{Humphrey Yang}, \bibinfo{person}{Zeyu Yan}, \bibinfo{person}{Nurcan Gecer~Ulu}, \bibinfo{person}{Ye Tao}, \bibinfo{person}{Jianzhe Gu}, \bibinfo{person}{Levent~Burak Kara}, {and} \bibinfo{person}{Lining Yao}.} \bibinfo{year}{2018}\natexlab{b}.
\newblock \showarticletitle{4DMesh: 4D Printing Morphing Non-Developable Mesh Surfaces}. In \bibinfo{booktitle}{\emph{Proceedings of the 31st Annual ACM Symposium on User Interface Software and Technology}} (Berlin, Germany) \emph{(\bibinfo{series}{UIST '18})}. \bibinfo{publisher}{Association for Computing Machinery}, \bibinfo{address}{New York, NY, USA}, \bibinfo{pages}{623–635}.
\newblock
\showISBNx{9781450359481}
\urldef\tempurl%
\url{https://doi.org/10.1145/3242587.3242625}
\showDOI{\tempurl}


\bibitem[\protect\citeauthoryear{Wen, Bae, and Rivera}{Wen et~al\mbox{.}}{2025}]%
        {wen2025recycling}
\bibfield{author}{\bibinfo{person}{Xin Wen}, \bibinfo{person}{S.~Sandra Bae}, {and} \bibinfo{person}{Michael~L. Rivera}.} \bibinfo{year}{2025}\natexlab{}.
\newblock \showarticletitle{Enabling Recycling of Multi-Material 3D Printed Objects through Computational Design and Disassembly by Dissolution}. In \bibinfo{booktitle}{\emph{Proceedings of the 2025 CHI Conference on Human Factors in Computing Systems}} \emph{(\bibinfo{series}{CHI '25})}. \bibinfo{publisher}{ACM}, \bibinfo{address}{New York, NY, USA}.
\newblock


\bibitem[\protect\citeauthoryear{Wu and Devendorf}{Wu and Devendorf}{2020}]%
        {Unfabricate}
\bibfield{author}{\bibinfo{person}{Shanel Wu} {and} \bibinfo{person}{Laura Devendorf}.} \bibinfo{year}{2020}\natexlab{}.
\newblock \showarticletitle{Unfabricate: Designing Smart Textiles for Disassembly}. In \bibinfo{booktitle}{\emph{Proceedings of the 2020 CHI Conference on Human Factors in Computing Systems}} (Honolulu, HI, USA) \emph{(\bibinfo{series}{CHI '20})}. \bibinfo{publisher}{Association for Computing Machinery}, \bibinfo{address}{New York, NY, USA}, \bibinfo{pages}{1–14}.
\newblock
\showISBNx{9781450367080}
\urldef\tempurl%
\url{https://doi.org/10.1145/3313831.3376227}
\showDOI{\tempurl}


\bibitem[\protect\citeauthoryear{Yan, Cheng, Lu, Lopes, and Peng}{Yan et~al\mbox{.}}{2023a}]%
        {yan2023future}
\bibfield{author}{\bibinfo{person}{Zeyu Yan}, \bibinfo{person}{Tingyu Cheng}, \bibinfo{person}{Jasmine Lu}, \bibinfo{person}{Pedro Lopes}, {and} \bibinfo{person}{Huaishu Peng}.} \bibinfo{year}{2023}\natexlab{a}.
\newblock \showarticletitle{Future Paradigms for Sustainable Making}. In \bibinfo{booktitle}{\emph{Adjunct Proceedings of the 36th Annual ACM Symposium on User Interface Software and Technology}} (San Francisco, CA, USA) \emph{(\bibinfo{series}{UIST '23 Adjunct})}. \bibinfo{publisher}{Association for Computing Machinery}, \bibinfo{address}{New York, NY, USA}, Article \bibinfo{articleno}{109}, \bibinfo{numpages}{3}~pages.
\newblock
\showISBNx{9798400700965}
\urldef\tempurl%
\url{https://doi.org/10.1145/3586182.3617433}
\showDOI{\tempurl}


\bibitem[\protect\citeauthoryear{Yan, Dhagude, and Peng}{Yan et~al\mbox{.}}{2025a}]%
        {MakeMaking}
\bibfield{author}{\bibinfo{person}{Zeyu Yan}, \bibinfo{person}{Mrunal Dhagude}, {and} \bibinfo{person}{Huaishu Peng}.} \bibinfo{year}{2025}\natexlab{a}.
\newblock \showarticletitle{Make Making Sustainable: Exploring Sustainability Practices, Challenges, and Opportunities in Making Activities}. In \bibinfo{booktitle}{\emph{Proceedings of the 2025 CHI Conference on Human Factors in Computing Systems}} \emph{(\bibinfo{series}{CHI '25})}. \bibinfo{publisher}{Association for Computing Machinery}, \bibinfo{address}{New York, NY, USA}.
\newblock
\urldef\tempurl%
\url{https://doi.org/10.1145/3706598.3713665}
\showDOI{\tempurl}


\bibitem[\protect\citeauthoryear{Yan, Lee, He, and Peng}{Yan et~al\mbox{.}}{2023b}]%
        {Magnetophoretic}
\bibfield{author}{\bibinfo{person}{Zeyu Yan}, \bibinfo{person}{Hsuanling Lee}, \bibinfo{person}{Liang He}, {and} \bibinfo{person}{Huaishu Peng}.} \bibinfo{year}{2023}\natexlab{b}.
\newblock \showarticletitle{3D Printing Magnetophoretic Displays}. In \bibinfo{booktitle}{\emph{Proceedings of the 36th Annual ACM Symposium on User Interface Software and Technology}} (San Francisco, CA, USA) \emph{(\bibinfo{series}{UIST '23})}. \bibinfo{publisher}{Association for Computing Machinery}, \bibinfo{address}{New York, NY, USA}, Article \bibinfo{articleno}{54}, \bibinfo{numpages}{12}~pages.
\newblock
\showISBNx{9798400701320}
\urldef\tempurl%
\url{https://doi.org/10.1145/3586183.3606804}
\showDOI{\tempurl}


\bibitem[\protect\citeauthoryear{Yan, Li, Zhang, and Peng}{Yan et~al\mbox{.}}{2024}]%
        {solderlesspcb}
\bibfield{author}{\bibinfo{person}{Zeyu Yan}, \bibinfo{person}{Jiasheng Li}, \bibinfo{person}{Zining Zhang}, {and} \bibinfo{person}{Huaishu Peng}.} \bibinfo{year}{2024}\natexlab{}.
\newblock \showarticletitle{SolderlessPCB: Reusing Electronic Components in PCB Prototyping through Detachable 3D Printed Housings}. In \bibinfo{booktitle}{\emph{Proceedings of the CHI Conference on Human Factors in Computing Systems}} \emph{(\bibinfo{series}{CHI '24})}. \bibinfo{publisher}{Association for Computing Machinery}, \bibinfo{address}{New York, NY, USA}, Article \bibinfo{articleno}{345}, \bibinfo{numpages}{17}~pages.
\newblock
\showISBNx{9798400703300}
\urldef\tempurl%
\url{https://doi.org/10.1145/3613904.3642765}
\showDOI{\tempurl}


\bibitem[\protect\citeauthoryear{Yan, Vartak, Li, Zhang, and Peng}{Yan et~al\mbox{.}}{2025b}]%
        {yan2025pcbrenewal}
\bibfield{author}{\bibinfo{person}{Zeyu Yan}, \bibinfo{person}{Advait Vartak}, \bibinfo{person}{Jiasheng Li}, \bibinfo{person}{Zining Zhang}, {and} \bibinfo{person}{Huaishu Peng}.} \bibinfo{year}{2025}\natexlab{b}.
\newblock \showarticletitle{PCB Renewal: Iterative Reuse of PCB Substrates for Sustainable Electronic Making}. In \bibinfo{booktitle}{\emph{Proceedings of the 2025 CHI Conference on Human Factors in Computing Systems}} (Yokohama, Japan) \emph{(\bibinfo{series}{CHI '25})}. \bibinfo{publisher}{Association for Computing Machinery}, \bibinfo{address}{New York, NY, USA}.
\newblock
\urldef\tempurl%
\url{https://doi.org/10.1145/3706598.3714276}
\showDOI{\tempurl}


\bibitem[\protect\citeauthoryear{Yedrissov, Khrustalev, Alekseev, Khrustaleva, and Vetrova}{Yedrissov et~al\mbox{.}}{2022}]%
        {yedrissov2022new}
\bibfield{author}{\bibinfo{person}{Azamat Yedrissov}, \bibinfo{person}{Dmitriy Khrustalev}, \bibinfo{person}{Alexander Alekseev}, \bibinfo{person}{Anastassiya Khrustaleva}, {and} \bibinfo{person}{Anastassiya Vetrova}.} \bibinfo{year}{2022}\natexlab{}.
\newblock \showarticletitle{New composite material for biodegradable electronics}.
\newblock \bibinfo{journal}{\emph{Materials Today: Proceedings}}  \bibinfo{volume}{49} (\bibinfo{year}{2022}), \bibinfo{pages}{2443--2448}.
\newblock


\bibitem[\protect\citeauthoryear{Zhang and Others}{Zhang and Others}{2023}]%
        {zhang2023recyclable}
\bibfield{author}{\bibinfo{person}{Firstname Zhang} {and} \bibinfo{person}{Others}.} \bibinfo{year}{2023}\natexlab{}.
\newblock \showarticletitle{vPCB: Recyclable Vitrimer-Based Circuit Boards}.
\newblock \bibinfo{journal}{\emph{ACM Transactions on Graphics}} \bibinfo{volume}{XX}, \bibinfo{number}{X} (\bibinfo{year}{2023}), \bibinfo{pages}{1--10}.
\newblock


\bibitem[\protect\citeauthoryear{Zhang, Biswal, Nandi, Frost, Smith, Nguyen, Patel, Vashisth, and Iyer}{Zhang et~al\mbox{.}}{2024}]%
        {zhang2024recyclable}
\bibfield{author}{\bibinfo{person}{Zhihan Zhang}, \bibinfo{person}{Agni~K Biswal}, \bibinfo{person}{Ankush Nandi}, \bibinfo{person}{Kali Frost}, \bibinfo{person}{Jake~A Smith}, \bibinfo{person}{Bichlien~H Nguyen}, \bibinfo{person}{Shwetak Patel}, \bibinfo{person}{Aniruddh Vashisth}, {and} \bibinfo{person}{Vikram Iyer}.} \bibinfo{year}{2024}\natexlab{}.
\newblock \showarticletitle{Recyclable vitrimer-based printed circuit boards for sustainable electronics}.
\newblock \bibinfo{journal}{\emph{Nature Sustainability}} \bibinfo{volume}{7}, \bibinfo{number}{5} (\bibinfo{date}{May} \bibinfo{year}{2024}), \bibinfo{pages}{616--627}.
\newblock
\showISSN{2398-9629}
\urldef\tempurl%
\url{https://doi.org/10.1038/s41893-024-01333-7}
\showDOI{\tempurl}


\bibitem[\protect\citeauthoryear{Zhu, Winagle, and Kao}{Zhu et~al\mbox{.}}{2024}]%
        {ecothread}
\bibfield{author}{\bibinfo{person}{Jingwen Zhu}, \bibinfo{person}{Lily Winagle}, {and} \bibinfo{person}{Hsin-Liu~(Cindy) Kao}.} \bibinfo{year}{2024}\natexlab{}.
\newblock \showarticletitle{EcoThreads: Prototyping Biodegradable E-textiles Through Thread-based Fabrication}. In \bibinfo{booktitle}{\emph{Proceedings of the CHI Conference on Human Factors in Computing Systems}} (Honolulu, HI, USA) \emph{(\bibinfo{series}{CHI '24})}. \bibinfo{publisher}{Association for Computing Machinery}, \bibinfo{address}{New York, NY, USA}, Article \bibinfo{articleno}{857}, \bibinfo{numpages}{17}~pages.
\newblock
\showISBNx{9798400703300}
\urldef\tempurl%
\url{https://doi.org/10.1145/3613904.3642718}
\showDOI{\tempurl}


\end{thebibliography}

\appendix 
\section{Appendix} \label{appendix}

This appendix expands on the workflow for creating circuit boards with \dissolvpcb by providing additional operational details that are not included in the main manuscript. It includes practical tips, recommendations, and precautions to support users throughout the process. 
An illustrated version of this guide is available in the accompanying GitHub repository (Footnote \ref{fn1}) as well.

\subsection{\dissolvpcb Production}
\begin{figure}[b]
    \centering
    \includegraphics[width = \columnwidth]{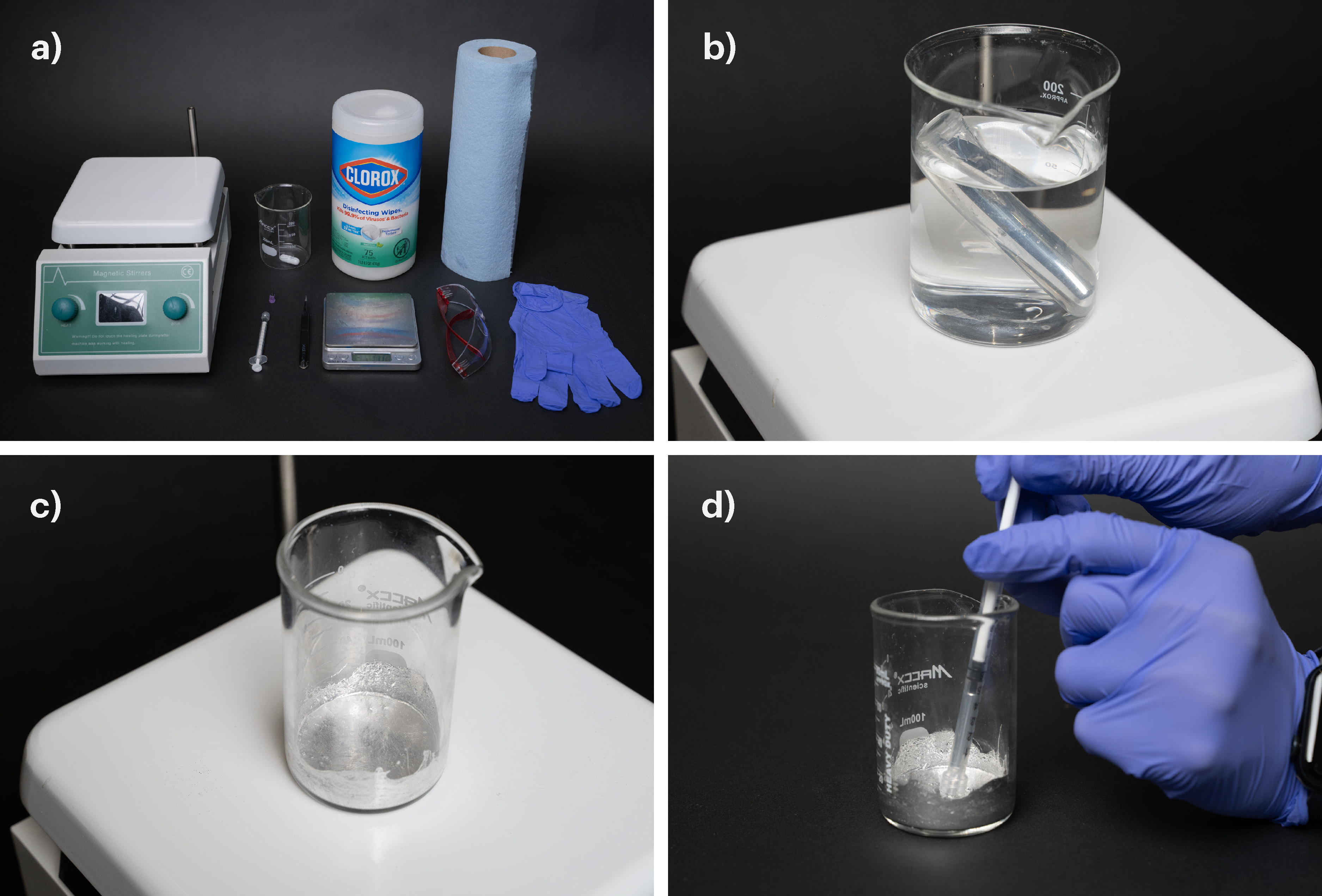}
    \caption{Recommended equipment and preparation of EGaIn: a) recommended equipment; b) melting gallium in a hot water bath; c) EGaIn after mixing; d) loading EGaIn into a syringe.}
    \label{fig:production}
\end{figure}

\paragraph{Tools \& materials.} 
Before producing \dissolvpcb, users are advised to have the following equipment (Figure \ref{fig:production}a): wet wipes (e.g., Clorox wipes) for cleaning EGaIn from surfaces; magnetic stirrers and a hotplate unit for preparing EGaIn; glass beakers; a scale; a syringe; and fine tweezers. 
Although EGaIn is non-toxic, personal protective equipment (PPE) is highly recommended. 
Wearing gloves, goggles, and a lab coat helps prevent contamination and minimize exposure to raw materials. 

\paragraph{PVA prints.}
After printing with PVA, it is recommended to purge any remaining PVA from the printer nozzle using PLA. 
If PVA clogs the nozzle, disassembling it from the printer and submerging it in water may help remove the remaining material. 

\paragraph{EGaIn preparation and storage.}
EGaIn is prepared by mixing gallium and indium in a 3:1 ratio by weight.
The preparation starts with warming the gallium in a hot water bath above \SI{30}{\degreeCelsius} to liquefy it (Figure~\ref{fig:production}b).
The liquefied gallium is poured into a glass beaker, followed by the addition of indium.
The mixture is stirred using a magnetic stirrer on a hot plate at a temperature above \SI{30}{\degreeCelsius} until the metals are fully combined into a uniform liquid metal (Figure \ref{fig:production}c).
To minimize oxidation, ensure the magnetic stir bar remains fully submerged during mixing.

Store EGaIn in an airtight container, or minimize exposure to air by covering it with a lid whenever possible.
If oxidation occurs, the viscosity can be fine-tuned by mixing in unoxidized EGaIn to achieve the desired consistency.

In most cases, EGaIn can be collected using a syringe or pipette (Figure \ref{fig:production}d). 
A beaker of NaOH solution is useful as a disposal container when working with EGaIn.

\paragraph{EGaIn injection.} 
In our implementation, we used a tapered steel Luer lock tip with a 23-gauge size for EGaIn injection.
When injecting EGaIn into printed channels, use minimal force and watch for potential backflow, especially in long or complex trace systems.
Excess EGaIn or spills can be cleaned using wet wipes and collected in a NaOH solution.
If an extrusion tip requires cleaning, submerge it in a NaOH solution and flush liquid through it to remove any residual EGaIn.

\paragraph{Components placement.}
When placing electronic components, be mindful of potential short circuits caused by excess EGaIn or poor contact resulting from insufficient EGaIn.
After securing the components with PVA glue, the curing process can be accelerated by drying them in a filament dryer or convection oven.

\subsection{\dissolvpcb Recycling}
\paragraph{Material \& component recovery.} 
The dissolution of the assemblies can be accelerated by using warm water and occasional stirring.
Once the board has fully dissolved, retrieve the components from the solution using tweezers.

\begin{figure}[b]
    \centering
    \includegraphics[width = \columnwidth]{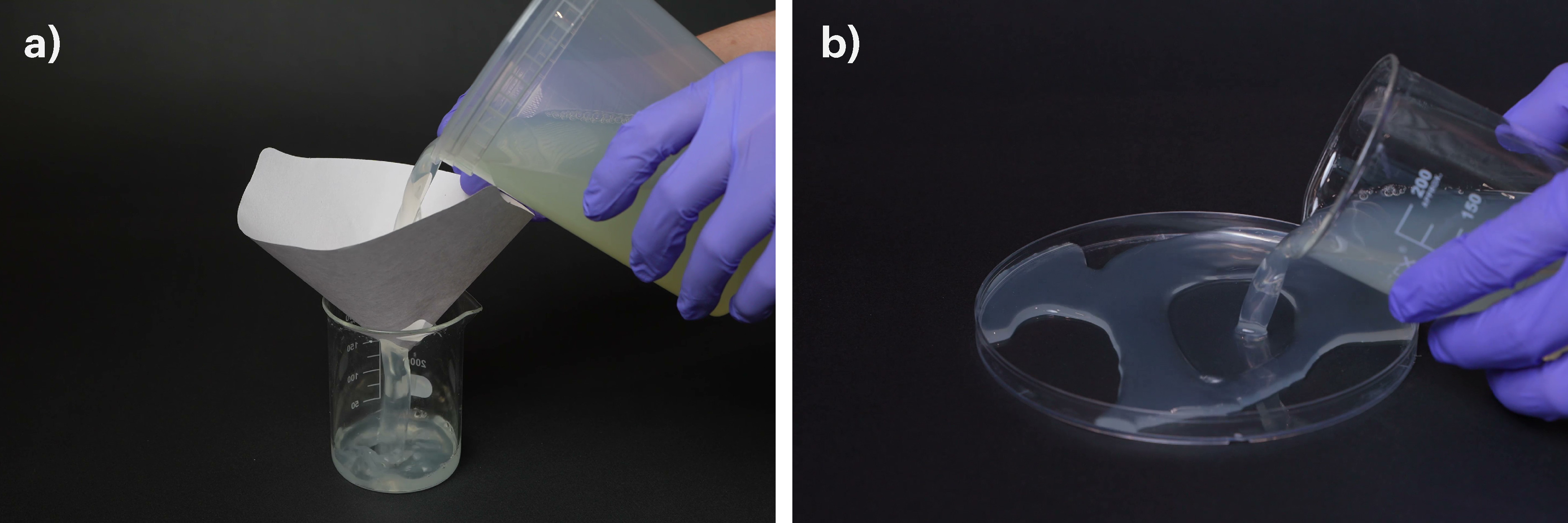}
    \caption{Processing PVA solution: a) filtering the recovered solution; b) pouring the filtered PVA solution into petri dishes for drying.}
    \label{fig:PVAsolution}
\end{figure}

\paragraph{PVA recycling.}
The PVA solution should be washed out of the container used for dissolving the assembly.
It should then be filtered to remove any contaminants before drying (Figure \ref{fig:PVAsolution}a).
Pour the filtered solution into petri dishes (Figure \ref{fig:PVAsolution}b) and allow it to dehydrate in open air, or use a hotplate to accelerate the process.
Dried PVA sheets can be rehydrated to form PVA glue or, alternatively, fully dried in a filament dryer and shredded for filament production.
The extrusion temperature should be high enough to plasticize the material without making it overly fluid.
We used \SI{185}{\degreeCelsius} for the extruder.
If moisture is present in the raw material, the extruded filament may contain air bubbles. In that case, repeat the shredding and extrusion steps to obtain high-quality filament.

\end{document}